\documentclass[12pt]{article}

\usepackage{epsfig}

\newcommand{\etal}{\emph{et al.\ }}
\newcommand{\pgf}{\noindent\hangindent=0.5truein\hangafter=1}
\newcommand{\qcrit}{$Q^*_D$\,}
\newcommand{\qshat}{$Q^*_S$\,}

\title{Catastrophic Disruptions Revisited}
\author{Willy Benz\\
        Physikalisches Institut\\
        University of Bern \\
        Switzerland \\
        email: wbenz@phim.unibe.ch \\
        and\\
        Erik Asphaug\\
        Earth Sciences Department\\
        University of California\\
        Santa Cruz, CA 95064 \\
        email: asphaug@cosmic.ucsc.edu}

\date{\small{To appear in Icarus}}

\begin{document}

\maketitle

\begin{abstract}
We use a smooth particle hydrodynamics method (SPH) to simulate
colliding rocky and icy bodies from cm-scale to hundreds of km in
diameter, in an effort to define self-consistently the threshold for
catastrophic disruption. Unlike previous efforts, this analysis
incorporates the combined effects of material strength (using a brittle
fragmentation model) and self-gravitation, thereby providing results in
the ``strength regime'' and the ``gravity regime'', and in between. In
each case, the structural properties of the largest remnant are
examined.

Our main result is that gravity plays a dominant role in determining the
outcome of collisions even involving relatively small targets. In the
size range considered here, the enhanced role of gravity is not due to
fracture prevention by gravitational compression, but rather to the
difficulty of the fragments to escape their mutual gravitational
attraction. Owing to the low efficiency of momentum transfer in
collisions, the velocity of larger fragments tends to be small, and more
energetic collisions are needed to disperse them.

We find that the weakest bodies in the solar system, as far as impact
disruption is concerned, are about 300 m in diameter. Beyond this size,
objects become more difficult to disperse even though they are still
easily shattered. Thus, larger remnants of collisions involving targets
larger than about 1 km in radius should essentially be self-gravitating
aggregates of smaller fragments. 

\end{abstract}

\section{Introduction}
\label{sec:intro}

Solar system evolution hinges on the outcome of collisions.  In order
for planetary accretion to proceed, for example, collisions must result
in larger, not smaller pieces. Asteroid dynamical families are in most
cases thought to be the outcome of catastrophically disruptive
collisions between parent bodies; smaller asteroids and interplanetary
dust may be the result of collisional cascades; asteroid and planetary
binaries (Ida and Dactyl, Pluto and Charon, Earth and Moon) may also be
an expression of impact.

Impacts or collisions can be grouped in three different categories
depending on outcome: cratering, shattering and dispersing. The first
category is defined by events leading to the formation of topographical
signatures (craters) accompanied by ejection of material but without
affecting the physical integrity of the main body. Shattering impacts,
on the other hand, are events that break the parent body into smaller
pieces.  Dispersing events are those which not only break the body into
pieces, but also manage to impart velocities to those fragments in
excess of escape velocity. Much observational evidence testifies to
these most energetic events: the dynamical asteroid families for example
(such as Koronis), and the iron meteorites which are fragments excavated
by impact from the cores of differentiated bodies.

It has become customary in the literature to characterize impacts in
terms of a specific energy threshold (the kinetic energy in the
collision divided by target mass). The threshold for a shattering event
is defined by \qshat, the specific energy required to break a body into
a spectrum of intact fragments, the largest one having exactly half the
mass of the original target. Dispersing events, on the other hand, are
defined by \qcrit, the specific energy required to disperse the targets
into a spectrum of individual but possibly reaccumulated objects, the
largest one having exactly half the mass of the original target. In the
strength regime, where gravity does not matter (fragments do not
reaccumulate), \qshat is obviously equal to \qcrit. In the gravity
regime however, \qcrit is always greater than \qshat, since the target
must be fragmented and also \emph{dispersed} by the event.

Laboratory experiments can be designed to determine this threshold for
small targets, i.e. targets in the strength dominated regime (see for
example Fujiwara \etal 1989; Davis \& Ryan 1990; Ryan \etal 1991). By
using up to meter-sized targets, Housen and Holsapple (1999) were able 
to confirm previous theoretical prediction of strength weakening with 
size. Impacts in pressurized targets (Housen 1991) designed to simulate 
self-gravitating bodies indicate that strength increases again in 
these ``larger'' targets.  However, these artificially pressurized 
targets are uniformly compressed while in truly self-gravitating 
bodies the overburden is a function of position. The interpretation 
of these experiments is therefore not completely straightforward.

The scales of planetary impacts are far different from what can be
studied directly in the laboratory, and extrapolations by more than a
dozen orders of magnitude in mass are required before reaching a range
relevant to asteroids and/or planetesimals. Detailed theoretical models
of disruptive impacts (Holsapple \& Housen 1986; Housen \& Holsapple
1990; Holsapple 1994) try to bridge this gap by establishing relations
among non-dimensional ratios involving impactor size, impact velocity,
target strength, density, etc. Such relations, deriving from dimensional
analysis, assume uniformity of process, structural continuity and other
idealizations, and cannot predict detailed outcomes.

Recent exponential increases in computational power have enabled
numerical simulations to become the method of choice to investigate
these issues in greater detail. Laboratory impact experiments are used
to validate the numerical models on small scales before extrapolations
to sizes relevant to solar system bodies are undertaken. However,
numerical attempts to determine \qcrit have for the moment been limited
to idealized geometry (two-dimensional axisymmetry) (Ryan \& Melosh
1998) or restricted to a subset of target sizes (gravity regime) (Melosh
\& Ryan 1997; Love \& Ahrens 1996) and low resolution (Love \& Ahrens
1996).

In this paper, we use a smooth particle hydrodynamics method (SPH) to
simulate colliding basalt and icy bodies from cm-scale to hundreds of km
in diameter, in an effort to define \emph{self-consistently} the
threshold for catastrophic disruption. Unlike previous efforts, this
analysis incorporates the combined effects of material strength (using a
brittle fragmentation model) and self-gravitation, thereby providing
results in the ``strength regime'' and the ``gravity regime'' and in
between. We begin with a short presentation of the physical model
(section \ref{sec:physics}) followed by a discussion of various
numerical issues pertaining  to our study (section \ref{sec:numerics}).
The results obtained from 480 different simulations are presented and
discussed in section \ref{sec:results}.

\section{Physical Model}
\label{sec:physics}

Our approach of dynamical fracture modeling has been described in detail
elsewhere (Benz and Asphaug 1994, 1995) and will only be summarized
here.

The equations describing an elastic solid are the usual conservation 
equations (mass, momentum and energy) in which the stress tensor has a 
non-diagonal part, the so-called deviatoric stress tensor. With the 
assumption that the stress deviator rate is proportional to the strain 
rate, i.e. Hooke's law, and a suitable equation of state (see section 
\ref{subsec:eos}) it is possible to numerically solve this set of 
coupled differential equations. Plasticity is introduced by suitably 
modifying the stresses beyond the elastic limit using a von Mises 
yielding relation. Not counting the equation of state, this approach 
requires 3 material-dependent constants: $\mu$ the shear modulus, 
$Y$ elastic limit and $E_{melt}$ the melt energy used to decrease the 
elastic limit with increasing temperature. Young's modulus can be 
computed from the knowledge of the bulk modulus (parameter $A$ in
section \ref{subsec:eos}) and the shear modulus according to $E=9 A \mu
/(3 A + \mu)$.

For the lower tensile stresses associated with brittle failure, we use a
fracture model based on the nucleation of incipient flaws whose number
density is given by a Weibull distribution (Weibull 1939; Jaeger \& Cook
1969)

\begin{equation}
   n(\epsilon) = k \epsilon^m
\label{eq:weibull}
\end{equation}

\noindent where $n$ is the number density of flaws having failure
strains lower than $\epsilon$. The Weibull parameters $m$ and $k$ are
material constants which can be determined from laboratory experiments
comparing failure stress to strain rate, although data are scarce. Table
\ref{tab:matcon} gives the numerical values of these material-dependent
constants used in this study. The Weibull parameters for basalt are a 
previously-determined (Benz \& Asphaug 1995) best match to the laboratory 
impact experiments of Nakamura \& Fujiwara (1991); parameters for ice 
are derived by fitting a line to the measurements of rate-dependent 
fracture stress published in Lange \& Ahrens (1983).

Once flaws are activated, fractures grow at constant velocity, about
half the sound speed. The extent to which fracture affects the local
properties of matter is described by a scalar state variable called
\emph{damage}. The possible values for $D$ range between $0$ (undamaged)
to $1$ (totally damaged). The ability to sustain shear or tension by
individual particles is reduced linearly with $D$ and vanishes for
$D=1$.  Because damage accrues according to the entire stress history of
a parcel of matter, only Lagrangian solutions to the hydrodynamics
equations are applicable.

Our model of dynamic fragmentation explicitly reproduces the growth of
cracks in a brittle elastic solid by rupturing bonds and forming new
free surfaces. Cracks grow when local failure strains are exceeded, and
stresses are relieved across the crack boundaries. The release of stress
along fracture walls increases differential stress at the crack tips,
driving cracks forward in the manner of an actual brittle solid. Tensile
and shear stresses are unsupported across disconnected regions, leading
to reduced average strength and sound speed (i.e. damage) in the body.
By producing actual cracks and fragments, our method at sufficiently
high resolution automatically takes into account friction between
fragments and bulking, effects which are included as recipes in
statistical damage models.

\subsection{Equation of State}
\label{subsec:eos}

We used the so-called Tillotson equation of state (Tillotson 1962)
which was specifically derived for high-velocity impact computations.
This equation has the advantage of being computationally expedient while
sophisticated enough to allow its application over a wide regime of
physical conditions.

For compressed regions and for cold expanded states where the energy
density ($E$) is less than the energy of incipient vaporization
($E_{\rm iv}$) the Tillotson eos has the form:

\begin{equation}
 P= \left[ a + \frac {b}{(E/(E_{0} \, \eta^{2}) + 1)} \right] \; \rho \, E
    + A \, \mu + B \, \mu^{2}
\label{eq:tillc}
\end{equation}

\noindent where $\eta=\rho/\rho_{0}$ and $\mu=\eta - 1$, such that $\rho$ 
is the
compressed density and $\rho_{0}$ is the zero-pressure density. Here
$a, b, A, B,$ and $E_{0}$ are referred to as the material-dependent
Tillotson parameters.

For expanded states, when the internal energy ($E$) is greater than the
energy of complete vaporization ($E_{\rm cv}$), the pressure has the form

\begin{equation}
 P= a \rho E + \left[ \frac {b \rho E} {(E/(E_{0} \, \eta^{2}) + 1)}
      + A \mu e^{- \beta \,  (\rho_{0} \, /\rho - 1)} \right] \;
      e^{ - \alpha \, (\rho_{0} \, /\rho - 1)^{2}}
\label{eq:tille}
\end{equation}

\noindent where $\alpha$ and $\beta$ are constants that control the
convergence rate of the equation to the perfect gas law. For
intermediate states, pressure is simply interpolated linearly between
both expanded and compressed states.

Tillotson parameters for a variety of geologic materials have been
compiled by Melosh (1989).  However, the most fundamental coefficients
(especially density and bulk modulus) for candidate materials are
typically different from those reported for specimens used in laboratory
impact fragmentation experiments, from which our Weibull coefficients
(Table 1) are derived.  We therefore make the following alteration to
the equations of state, in an effort to best characterize the material
in the fragmentation experiments:  We substitute the measured density
for the published Tillotson reference density of the most similar
material, and the measured bulk modulus for the Tillotson parameter
$A$.  Because the Tillotson parameter $B$ is not known for the specific
target material, we make use of the fact that $B \simeq A$ for most
geologic materials.  This is probably a good assumption for basalt but
may underestimate the second-order pressure terms in ice, making it more
``rock-like".  We summarize in Table \ref{tab:tillotson} the Tillotson
parameters used in this paper.  This situation is not ideal.  However,
given the rudimentary understanding of asteroidal and cometary
compositions (they are surely neither pure basalt nor pure water ice!),
we feel that there is little to be gained until comprehensie
experimental work, combining dynamic strength and equation of state, can
be performed on more representative candidate materials at the
appropriate temperature.

\section{Numerical Issues}
\label{sec:numerics}

In this section we discuss various issues related to the numerical
methods used to either simulate the impacts or analyze the results
of the simulations.

\subsection{Impact Simulations}
\label{subsec:simul}

Fracture depends on the entire stress history of a given piece of
material. A Lagrangian approach, in which the frame of reference is
attached to the material, is therefore the natural framework for solving
the equations briefly described in section \ref{sec:physics}; Eulerian
codes to date cannot accurately follow stress history and the
development of cracks. Conventional Lagrangian codes, however, are
unable to handle large material deformations, as tangling and
deformation of the grid severely affect the accuracy of derivatives.

Smooth Particle Hydrodynamics (see reviews by Benz 1990; Monaghan 1992)
does not suffer from this problem. We have developed a three-dimensional
SPH code capable of simulating dynamical fracture of brittle material
(Benz \& Asphaug 1994, 1995). Our SPH code being an explicit code, the
size of the time step is limited by the Courant condition. In practice,
this means that the time step cannot exceed a fraction of the time
needed by an elastic wave to cross a resolution element. If this element
is of size $h$, numerical stability requires that $dt < h / c$, where
$c$ is the wave speed. With 42,000 particles, $h$ is of order $R/35$
where $R$ is the target radius. As an estimate for wave speed we take
$c=\sqrt{A/\rho}$ where $A$ is the bulk modulus. Taking the values
appropriate for basalt, we get $c=3.2$ km/s which leads to an upper
bound for the time step given by $dt < 9 \times 10^{-3} \, R$, where $dt$
is in seconds and $R$ in km.

Gravitational clumping of fragments and the formation of rubble piles on
the other hand takes place on a dynamical timescale given by
$\tau=\sqrt{3\pi/16G\rho}$. For basalt, we obtain a value for the
clumping timescale of $\tau=1.8 \times 10^3$ s. The number of timesteps
required to follow a collision up to the time where gravitational
reaccumulation occurs is therefore $N > \tau/dt = 2 \times 10^{10} / R$.
For targets of radius 100m, this requires a number of timesteps greater
than $10^6$, which is currently prohibitive.

While these two vaslty different timescales prevent us from simulating
the entire process directly, they also mean that self-gravitation is
entirely decoupled from the impact phase for intermediate targets (see
Asphaug 1997). Clumping can therefore be studied independently as
part of the postprocessing analysis of the impact simulations proper
(see section \ref{subsec:frag}). This has also the additional advantage
not having to solve Poisson's equation (self-gravity) during the impact
phase. We therefore use a faster linked list in rank space algorithm to find
the neighbor particles rather than a hierarchical tree. While the
dynamical effects of self-gravity are not included in the simulations,
the fracture shielding effect due to compression is included in the
code(see section \ref{subsec:flaw}).

The code's behavior in the elastic regime was extensively tested against
simple analytical solutions, while the fracture modeling was tested by
simulating laboratory impact experiments (Nakamura and Fujiwara 1991;
Ahrens and Rubin 1993) leading to core-type and radial fragmentation and
extensive spallation. The code was able to reproduce the laboratory
experiments to a level of detailed accuracy never achieved before,
including shape, ejection speed and rotation of fragments, trajectories
of far-field fracture, and post-fragmentation sound speed relationships.

\subsection{Flaw Assignments}
\label{subsec:flaw}

We use the explicit flaw assignment procedure described in Benz \&
Asphaug (1995) using the material-dependent Weibull parameters listed in
Table \ref{tab:matcon}.  Our explicit flaw method improves upon the
mixed explicit-implicit method of Benz \& Asphaug (1994).  These methods
are wholly independent of numerical resolution (in terms of the assigned
flaw statistics) and lead to rate- and size-dependent fracture
thresholds.

Each particle is assigned a number of discrete fracture thresholds
distributed at random from the  underlying Weibull distribution (Eq. 1).
When local stresses exceed such a threshold, damage $D$ is allowed to
grow until the local stresses decrease again, or until $D$ reaches its
maximum authorized value. This maximum authorized value for particle $i$
is given by $D^i_{\rm max}= N^i_a/N^i_{\rm max}$ where $N^i_a$ is the
currently active number of flaws and $N^i_{\rm max}$ the number of
fracture thresholds assigned to particle $i$.  This particular algorithm
of damage accrual and authorization ensures that flaw distribution and
activation are not linked to numerical resolution.

For fracture to proceed, stresses must first overcome the local
gravitational overburden. To include what amounts to an effective
\emph{gravitational strengthening} without having to solve the full
Poisson equation during the simulations, we use the technique (Asphaug
and Melosh 1993) of adding to the local stress tensor (determined from solving
the elasto-dynamical equations) an isotropic lithostatically
equilibrated stress ($P_l(r)= \frac{2 \pi}{3} G \rho^2 (R^2-r^2)$).
The sum, the ``total fracture stress", is converted into a Weibull 
strain for use
in Eq. 1 by dividing the largest tensile principal stress by the
modified Young's modulus, as pioneered by Melosh \etal (1992). Note that
the difference between the rock fracture stress and the total fracture
stress is negligible for targets smaller than a few 10's of km radius, and
that for larger targets, strengthening towards the interior may be due as
much to thermodynamical processes (annealing) than to gravity per se.

\subsection{Fragment search}
\label{subsec:frag}

As outlined above, late time gravitational evolution can for
intermediate size targets be decoupled from the impact physics itself.
Thus, we perform the characterization of the collisional outcome as a
postprocessing step. Our approach to find the final collisional outcome
proceeds in two steps.  First, and regardless of target size, we begin
by searching for intact (undamaged) fragments, i.e. fragments that are
held together by material strength alone. Our method makes no
assumptions regarding their number, geometry, or location, as they are a
natural outcome of the fracture trajectories resulting from a given
simulation. Contrary to statistical fragment approaches, our fragments
are explicitly defined by the network of cracks resulting from the
impact. We use a friends-of-friends algorithm to identify individual
monolithic fragments defined as contiguous regions of particles held
together by strength and surrounded by strengthless or empty regions. At
the end of this procedure, we obtain for each fragment its mass,
position, velocity, angular momentum and moment of inertia.  This is the
``shattering" spectrum of fragment sizes discussed in the Introduction.

As soon as the target size exceeds 50-100m, searching for intact
fragments is no longer sufficient, as some of them may be able to
reaccumulate due to gravity, leading to remnants incorporating multiple
fragments. We search for these gravitationally-bound aggregates by
applying a well known iterative procedure adapted from the techniques
used in simulations of galaxy formation. The procedure starts by
computing the binding energy of all particles and/or fragments with
respect to the largest fragment, or if too small, the particle closest
to the potential minimum. This serves as the seed for nucleating the
total bound mass.  Unbound particles are discarded, and the center of
mass position and velocity of the aggregate is computed. The binding
energy of all the remaining fragments and particles with respect to this
new position and velocity is again computed. Unbound particles are again
discarded and the procedure iterated until no particles are discarded.
Typically, convergence is achieved after a few iterations ($\approx
5-10$) with only very few particles being lost after the first 2-3
steps. Finally, we check that particle/fragment members of this
gravitationally bound aggregate are also close spatially, using again a
friends-of-friends algorithm. Mass, position, velocity, angular momentum
and moment of inertia are also determined for this gravitationally bound
aggregate, which can be made of a collection of smaller fragments and/or
individual particles. Because of the limited number of particles used,
we limit our search to the single largest aggregate and do not attempt
to search for smaller ones.

The algorithm has been tested extensively in both the strength and
gravity dominated regime by comparing end results of simulations with
predictions made at early times by the cluster finding algorithm. To
overcome the time step problem in the gravity dominated regime, purely
hydrodynamical simulations were used to allow the code to compute to
sufficiently late times.

For each simulation, we therefore identify and characterize the largest
object present at the end of the collision. This object may either be an
intact fragment (in the strength dominated regime) or a gravitationally
bound aggregate of fragments (gravity dominated regime). In the latter
case, we also search for the largest intact fragment belonging to the
aggregate (see Fig. \ref{fig:impcart}), e.g. the largest component of
the resulting ``rubble-pile''.

\subsection{Initial Conditions}
\label{subsec:initial}

We have considered 8 different target radii: 3cm, 3m, 300m, 1km, 3km,
10km, 30km, and 100km, and two different material types: basalt and ice.
For each target we have computed impacts for five angles of incidence
measured from the surface normal: 0, 30, 45, 60, and 75 degrees. For
each material we have considered two impact velocities: for icy targets
0.5 km/s and 3 km/s; for basalt targets 3 km/s and 5 km/s. In each case,
\qcrit  was computed from a parabolic fit to three different simulations
in which only the impactor mass was modified. The entire study therefore
represents a total of 480 different simulations which were automatically
handled by special ``driver" software.  As each simulation can take up
to a few days of high performance workstation time, this represents a
significant body of computational effort.

Because of the statistical nature of this study, we had to limit
ourselves to a relatively small number of particles. In all cases, we
used $42,000$ particles to model the target. This number was found in
convergence-test comparisons between numerical simulations and
laboratory impact experiments (Benz \& Asphaug 1994, 1995) to be
sufficient to determine reliably the characteristics (size, velocity and
rotation) of the largest fragment. Thus in all what follows, we shall
concentrate only on the characteristics of the largest fragment.

The projectile was modeled using $800$ particles for the $3$ km/s and
$5$ km/s impacts and $7,000$ for $0.5$ km/s impacts since in this case
the impactor was much larger. (In fact, in the case of $100$ km large
icy targets and large incidence angles the required projectile for a
given value of $Q$ was sometimes bigger than the target!) The
simulations were carried out in time until no further significant
changes occurred in the extent of damage endured by the target and in
the velocity of the ejected particles.

\section{Results}
\label{sec:results}

\subsection{Catastrophic disruption threshold \qcrit}
\label{subsec:qcrit}

For a given impact geometry, velocity and target (parent body) material,
\qcrit is determined by interpolation between three different
simulations spanning a range of incident kinetic energy per unit target
mass ($Q$) chosen to yield largest remnant masses ($f_{\rm max}=M_{\rm
lr}/M_{\rm pb}$) generally in the range $0.3-0.7$. In the expression
above, $M_{\rm lr}$ represents the mass of the largest remnant
(including gravitational reaccumulation if applicable) and $M_{\rm pb}$
the mass of the parent body (or target). A parabolic fit ($f_{\rm max}=
a\,Q^2 + b\,Q + c$) to these results is computed and \qcrit determined
by solving for $f_{\rm max}$(\qcrit)$=0.5$. The results of these
calculations are shown in Figs. \ref{fig:qbas3kms}, \ref{fig:qbas5kms},
\ref{fig:qice3kms} and \ref{fig:qice05kms} for the various velocities
and material type. In these figures, each dot represents the value of
\qcrit determined from the parabolic fit. The dots corresponding to the
same angle of incidence of the projectile are connected by a solid line.

We recover in these figures the well known functional dependency of
\qcrit with target size, namely that \qcrit decreases with size for
small targets while it increases with size for larger targets. These 
two behaviors correspond to collisions occurring in a strength or 
gravity dominated regime. The transition between the two regimes occurs 
in the range $100 {\rm m} \leq R \leq 1 {\rm km}$ for both ice and 
basalt. We note that for a given target size, \qcrit is a strong 
function of the projectile's angle of incidence.  Differences in 
\qcrit between a head-on and a $75$ degree impact reach about a 
factor 10.

The increase of \qcrit in the gravitational regime is due to two
factors. The most important is the fact that even though bodies
exceeding 1km in radius are almost entirely shattered (see section
\ref{subsec:largestf}) in a \qcrit collision, the pieces do not all
disperse because their relative velocities are smaller than the escape
velocity of the aggregate. Hence, for catastrophically disrupted and
dispersed bodies larger than $\approx 1$ km, all the largest remnants
are found to be gravitationally bound aggregates of smaller fragments.
The second effect is due to gravitational shielding of the central
region of the target (see section \ref{subsec:flaw}) but as we shall see
in section \ref{subsec:largestf} this affects only the largest targets
in our mass range.

In many studies, one is not interested in the outcome of a particular
collision but rather in the collisional evolution of an entire
population. For these statistical studies, we can compute an effective
threshold which is independent of angle of incidence, namely
\qcrit$(\alpha)$ averaged over all possible impact geometries but at
fixed relative velocity. For an isotropic distribution of incoming
projectiles (at infinity), Shoemaker (1962) showed that the probability
distribution for impacting with an angle between $\alpha$ and $\alpha +
d\alpha$ is given by

\begin{equation}
   P(\alpha) d\alpha = 2 \sin(\alpha) \cos(\alpha) d\alpha \ \ \ \ \
   0 < \alpha < \pi/2
\label{eq:prob}
\end{equation}

\noindent regardless of a planet's mass.  Using this probability 
distribution, we 
define a mean catastrophic
disruption threshold $\bar {Q}^*_D$ by

\begin{equation}
  \bar{Q}^*_D = \int_0^{\pi/2} 2 Q^*_D(\alpha) \sin(\alpha) \cos(\alpha)
              d\alpha
\label{eq:qmean}
\end{equation}

\noindent This integration is carried out numerically with a simple
trapezoidal method and using the \qcrit$(\alpha)$ determined from the
simulations. The results are displayed in Fig. \ref{fig:qbasmean} for
basalt targets and in Fig. \ref{fig:qicemean} for icy targets.

In order to compare with disruption thresholds published in the
literature and to allow these results to be used by others, we fitted
(by eye) an analytical curve to $\bar{Q}^*_D$ of the following
functional form

\begin{equation}
   \bar{Q}^*_D = Q_0 \left( \frac{R_{pb}}{1 \rm cm} \right)^a + B \rho
               \left(\frac{R_{pb}}{1 \rm cm}\right)^b
\label{eq:qfr}
\end{equation}

\noindent where $R_{pb}$ is the radius of the parent body (or target),
$\rho$ the density of the parent body (in g/cm$^3$)and $Q_0,B,a,b$
constants to be determined. This functional form is often encountered in
scaling law approaches with the two terms representing the two distinct
physical regimes dominating the dynamics: 1) material strength (first
term on the right, with $a<0$) and 2) self-gravity (second term on the
right, with $b>0$). The values obtained for the coefficients are listed
in Table \ref{tab:qbarcon} and the fits are represented as lines on
Figs. \ref{fig:qbasmean} and \ref{fig:qicemean}.

Note that the slopes in the gravity regime ($b$) are somewhat different
between basalt and ice, and even for ice the two slopes corresponding to
the two velocities differ slightly. This could be related to the fact
that for equal mass targets, ice material is lifted from an initially
higher potential (less negative potential energy) corresponding to the
larger equal-mass target diameter. Alternatively, because the shock
imparts different velocities to fragments according to different
material characteristics (the velocity field determines the subsequent
gravitational reaccumulation), equation of state distinctions sensitive
to both material type and impact speed may be the culprit.

Also note that in the gravity regime, faster impacts are less disruptive
than slower impacts of the same energy, for both rock and ice.  This is
due to the greater efficiency of momentum coupling for slower impacts. 
The governing factor for gravity regime disruption is not shattering,
but motion towards gravitational separation.  For basalt, this trend
continues into the strength regime, although that is coincidental, since
the strength regime depends on entirely different aspects of collisional
physics (flaw activation).  For ice, the opposite is true: slower
impacts (of the same energy) result in less disruption. Evidently ice is
easier to fracture at high strain rates than at low strain rates,
relative to basalt.  This is either because it has more flaws available
at low activation energies (consistent with its Weibull distribution),
or because vaporization at high impact speed may contribute to
fragmentation, which is not the case for the subsonic (500 m/s)
collisions.

Now we compare our averaged dispersion threshold with other published values
in Fig. \ref{fig:compq}. In this figure, we reproduced the
\emph{disruption} thresholds obtained either from scaling laws or
numerical simulations in the strength or gravitational dominated regime.
For small targets, our estimate of the threshold agrees well with the
determination of Holsapple (1994). The largest differences occur for
large targets for which we predict that they are significantly stronger
(more difficult to disrupt and disperse) than previously estimated. This
does not arise due to a significant change of slope (notice our slope is
close to the one predicted by Holsapple 1994 or Melosh \& Ryan 1997) but
because the turn over from strength to gravity dominated targets occurs
at smaller sizes.

We also display in Fig. \ref{fig:compq} the recent determination of 
\qcrit by Durda \etal (1998). Their curve is determined by requiring
that numerical models of the collisional evolution of the main belt 
asteroids fit the observed size distribution of these objects. 
Interestingly, they obtain that objects of order 100-200m in diameter
are the weakest objects, a conclusion confirmed by our simulations.
However, besides the agreement on the size of the weakest objects, our
results (as well as all other determinations of \qcrit), differ from 
the results of Durda \etal by an order of magnitude or more. The origin
of this discrepancy is not clear. On one hand, we note that the values
of \qcrit determined by Durda \etal are not obtained from a simple fit 
to observed sizes but assume an underlying collisional model which 
might not be predicting accurately collisional outcomes. On the other 
hand, it may also be possible that asteroidal material has very different 
mechanical properties than the material tested in the laboratory. 

As already noticed by many (most recently by Ryan and Melosh 1998), the
efficiency at which the kinetic energy is transmitted from the impactor
to individual fragments is extremely low. Since the largest remnants in
collisions involving targets larger than 1km are gravitationally bound
rubble piles (see section \ref{subsec:largestf}) their mass is
determined ultimately by the velocity distribution imparted to the
fragments during the impact. A fragment will remain bound if its
velocity remains below the escape velocity of the aggregate of all slow
moving fragments. This explains why targets as small as 1km radius are
already significantly strengthened by gravity against dispersal.

For a given material type, the radius of the weakest object $R_{\rm
weak}$, is obtained by finding the radius for which $d\bar{Q}^*_D/dR=0$.
From equation \ref{eq:qfr} the value of the radius of the weakest object
is given by

\begin{equation}
   R_{\rm weak} = \left( \frac{-B \rho b}{a Q_0} \right)^{\frac{1}{a-b}}
\label{eq:rweak}
\end{equation}

\noindent Table \ref{tab:rweak} gives the values derived for $R_{\rm weak}$
using the values of the parameters listed in Table \ref{tab:qbarcon}.  For 
both materials and impact speeds, $R_{\rm weak} \simeq$ 100--200 m.

These values are smaller than those derived
in other studies. For example, Holsapple (1994) based on scaling laws
gives 3 km as the transition point between the two regimes. Melosh \& Ryan
(1997) as well as Love \& Ahrens (1996) from numerical simulations give
numbers in the range 200-400m.

\subsection{Largest remnants}
\label{subsec:largestrem}

While a collision occurring at \qcrit leaves by definition a largest
remnant with mass equal to half the target mass, collisions occurring
with different $Q$ leave remnants of different masses. We can therefore
use all our simulations (whose intent was to bracket \qcrit) to investigate 
how the mass of the largest
remnant depends upon the incident kinetic energy per gram of target
material ($Q$).

Figs. \ref{fig:icemfq} and \ref{fig:basmfq} show the dependency of
the mass of the largest fragment on the impact energy obtained in our
simulations. To facilitate the interpretation of these results, the
impact kinetic energy per gram of target material has been normalized
to \qcrit for each target size and projectile angle of incidence. All
simulations (120) involving a given material type and impact velocity
have been plotted on the same plot. The different symbols correspond
to targets of different sizes.

These figures clearly show that when normalized to \qcrit the relative
mass of the largest remnant is a well-defined, simple function of $Q$ and
is independent of target size and/or angle of incidence. The dependence
upon these parameters enters only in \qcrit! The increased scatter
in the points for small mass fragments is probably due to the inherent
numerical difficulties in resolving these smaller objects (at fixed
resolution).

This relationship is remarkable since the mass of the largest remnant
is not determined by a single process, but either by material strength
or gravity depending on target size! In this regard it is interesting
to note that in the case of low velocity collisions on icy targets the
correlation is significantly worse, especially for large targets ($R
\ge 3$ km). It is unclear why this is the case; however, in these
cases it is worth pointing out that, due to the low velocity, the
projectile is sometimes as big as the target and that the relative 
velocity is significantly smaller than sound speed, indicating that 
there might be a different disruption regime at low velocity.

Except for the case discussed above, we find that the relative mass of
the largest remnant can be well represented by the following expression
and that this single expression holds for targets ranging from 3cm to 
100km and angles of incidence between 0 and 75 degrees.

\begin{equation}
   \frac{M_{\rm lr}}{M_{\rm pb}} = - s \left( \frac{Q}{Q^*} - 1 \right)
                                  + 0.5
\label{eq:mfq}
\end{equation}

\noindent These lines are drawn on the various figures and their slopes 
are remarkably similar: for basalt $s=0.35$ for $v=5$km/s, $s=0.5$
for $v=3$km/s and for ice $s=0.6$ for $v=3$km/s. The case for ice
at $v=0.5$ km/s for some still unknown reason does not yield such a 
tight relation. We have not attempted a real fit to the data and only 
plot the line derived for the $v=3$ km/s.

\subsection{Largest intact fragments}
\label{subsec:largestf}

We now determine the largest \emph{intact} (unshattered) fragment,
$M^*_{\rm lif}$, in collisions occurring at \qcrit. Here by undamaged we
mean a fragment for which material strength still plays an important
role in the cohesive properties of the object even though the fragment
may no longer have its original strength.  We are looking for the
largest boulder in the final rubble-pile.

For each target radius and projectile angle of incidence, we obtain from
the three simulations performed a parabolic relation $M_{\rm lif}=f(Q)$.
We determine $M^*_{\rm lif}$ by setting $M^*_{\rm lif}=f(Q^*_D)$ where
\qcrit has been determined using the parabolic fit described in section
\ref{subsec:qcrit}. The values obtained for the mass of these largest
intact fragments are displayed in Figs. \ref{fig:fraginice} and
\ref{fig:fraginbas} for both material types and impact velocities.

Not surprisingly, for small targets ($R \leq 100$m) the largest undamaged
fragment is equal to half the original target mass regardless of impact
parameter. This is simply because in the strength dominated regime the
mass of the largest intact fragment is equal to half the target mass by
definition of \qcrit. However, for targets larger than $300$ m and
regardless of material type and/or impact velocity, the mass of the
largest intact fragment drops rapidly even though the collisions
occurred at \qcrit. This reflects the fact that collisions involving
parent bodies this size and larger take place in the gravity dominated
regime. This regime is therefore characterized by the fact that the
largest remnant is not an intact fragment but a gravitationally bound
aggregate of fragments. Due to the precipitous nature of these curves
for target radii larger than a few 100 m, we dont expect any monolithic
object of this size. On the other hand, we expect a wide variety of
internal structures for objects in the 30 m - 300 m range.

For the largest targets considered in this study we notice a marginal
trend for $M^*_{\rm lif}$ to rise again. This effect is due to
gravitational strengthening of the target discussed in section
\ref{subsec:flaw}. However, we stress that this effect is only of
moderate importance: the main role of gravity in this size range is to
allow for the formation of gravitationally bound rubble piles.
Gravitational strengthening appears also to be strongly dependent on
initial impact parameter. For example, in the case of an icy target of
100km radius a grazing impact occurring at \qcrit and with an incidence
angle of $\alpha=75^\circ$ leaves behind a largest undamaged fragment of
$\approx 0.015M_{pb}$ while at $\alpha=0^\circ$ the same fragment is
smaller than $0.003 M_{pb}$.

Final objects in the range $1{\rm km} \leq R_{pb} \leq 100{\rm km}$ are
found to be essentially gravitationally bound aggregates of smaller
fragments. Whether or not observed asteroids and comets in this size
range are indeed rubble piles depends upon whether they have suffered a
\qcrit collision during their history. Further complications exist; for
example, Asphaug {\it et al.} (1998) have shown that target geometry
(shape) and internal structure (pre-fracture) can significantly
influence collisional outcome. The spherical homogeneous intact solids
considered here may be idealized.  A ``contact binary" asteroid may for
instance suffer catastrophic disruption of its impacted lobe, with
little or no disruption occurring on its unimpacted lobe.  A target
which is {\it already} a rubble-pile may similarly be more difficult to
disperse by impact, due to the inefficient coupling of impact energy.

\subsection{Ejection velocity of largest remnants}
\label{subsec:vejec}

For each target size, material type and impact parameter we determined
the ejection velocity of the largest mass remnant in each of the three
simulations bracketing \qcrit. We note that this velocity is not related
in a straightforward manner to the usual energy partitioning
coefficient, $f_{\rm KE}$, namely the fraction of kinetic energy going
into fragment kinetic energy. Given that the kinetic energy is
\emph{not} distributed uniformly over all fragments, but rather carried
away by a small amount of mass moving very fast, we believe the
coefficient $f_{\rm KE}$ to be of little use to address the dynamics of
the largest fragments. We therefore focus our attention on determining
the actual velocity of the largest fragment or aggregate in each
simulated collision, numerical resolution preventing us from studying
the smaller ones.

These ejection velocities as a function of normalized fragment mass are
displayed in Figs.(\ref{fig:vbas5kms}, \ref{fig:vbas3kms},
\ref{fig:vice3kms}, \ref{fig:vice05kms}) for both material types and
collision velocities.  Velocity is measured relative to the center of
mass of the original target.

In each figure, the upper panel shows the actual ejection velocity. 
Each different symbol corresponds to a different initial target size 
regardless of angle of incidence. The lower panel shows the ejection 
velocity normalized to the target's escape velocity for initial targets 
greater that 1km. 

We note that for a given parent body size, larger remnants have lower
ejection velocities regardless of angle of incidence. In fact, it is
remarkable how little influence the impact parameter seems to have on
the ejection velocity of the largest remnant. For each target size, the
ejection velocity of the largest remnant is to a good approximation a
simple decreasing linear function of its fractional mass (in the domain
$0.15 \le M_{lr}/M_{pb} \le 0.8$). In addition, we note that for the
largest fragments, i.e. the one for which gravity is the dominant
cohesion force, we can almost remove the target size dependence by 
normalizing the ejection velocity by the parent body's escape velocity. 
In other words, the outcome velocity of the largest fragment normalized 
to initial target escape velocity is (within some considerable scatter)
independent of target size and impact parameter. The fact that the 
velocity of the largest remnant is a relatively constant fraction of
the target's escape velocity is probably due to the a priori requirement
that in the gravitational regime about half of the initial mass 
must escape. 

In order to gain insight regarding which collisions lead to the fastest
moving largest remnants, we have analyzed how the ejection velocities
depend upon the ratio of impactor to target size, $R_{i}/R_{pb}$. We
stress again that because of numerical resolution, we are able to
analyze only the largest remnants and not the entire ejecta
distribution. Thus, it is unrealistic from our data to determine the
actual kinetic energy transfer efficiency $f_{\rm KE}$. 

Figs. \ref{fig:basvr} and \ref{fig:icevr} display the ejection velocity
of the largest remnants as a function of $R_{i}/R_{pb}$ regardless of
angle of incidence. The two different symbols correspond in each cases
to the two different impact velocities. Apparent from these figures is
the fact that the ejection velocity rises with $R_{i}/R_{pb}$ in a
monotonic fashion. The ``width'' of the curve is mainly determined not
by scatter but by the relation between ejection velocity and remnant
mass (see section(\ref{subsec:vejec}). Thus, regardless of angle of
incidence, collisions will give rise to fast moving remnants if the size
of the impactor becomes comparable to the size of the parent body.

In regard to the collisional origin of asteroid families, we note that
velocities of order 100 m/s are easily obtained for basaltic targets
provided the impactor size is at least about half the parent body size.

\section{Conclusions}

We have presented a self-consistent three-dimensional treatment of impact
disruption leading from the laboratory and the strength regime, where our
SPH code has been exhaustively calibrated and tested, all the way out to the
gravity regime collisions responsible for the formation of asteroid families
and planetary accumulation.  While some parameters (such as shape and
pre-fragmentation and rotation) have yet to be fully explored (see for
instance Asphaug {\it et al.} 1998) the 480 runs summarized here provide a
robust constraint on the outcome of catastrophic collisions. 

In particular, we have demonstrated that bodies ~100 - 200 m 
radius (depending on impact speed and composition; see Table 4)
are the weakest objects in the solar system, with all bodies this size and 
larger being dominated by self-gravitational forces, rather than material 
strength, with regard to impact disruption. This enhanced role of gravity 
is not due as usually assumed to fracture prevention by gravitational 
compression. It is due to the difficulty of the fragments to escape their 
mutual gravitational attraction. Owing to the generally low efficiency of 
momentum transfer in collisions, the velocity of larger fragments tends 
to be small, and more energetic collisions are needed to disperse them. 
Remarkably, the efficiency of momentum transfer (while still small) is 
found to be larger for larger projectiles. Thus, at a fixed collisional 
energy, a low velocity high mass projectile will lead to a higher fragment 
velocity that a small mass high speed projectile.

This increased role of gravity implies that the threshold for disruption 
is actually significantly larger than previously assumed. The upshot of 
this is that any catastrophic collisions leading to disruption \emph{must}
occur at an energy far exceeding the threshold for shattering the parent
body. These necessarily high strain rate collisions imply by the nature of 
the fracturing process that many cracks must grow to release the stresses, 
preventing any sizeable fragment from surviving. Thus, catastrophic collisions 
of this nature can only result in the formation of gravitationally bound 
aggregates of smaller fragments.

We shall continue to examine the outcome of these simulations for
information regarding angular momentum transfer during impact, and for
anticipated cumulate structures for large fragments from large parent
bodies, motivated by the possibility of re-creating the events which led
to the formation of known asteroid families (Benz and Michel, in preparation).  
As available computing power permits, a more detailed parameteric
exploration (varying Weibull coefficients, shape and internal structure)
is someday hoped for.  As it stands, our chosen materials (basalt and
ice) represent broad-based choices, and the fact that both give similar
results implies that catastrophic disruption is perhaps not very
material dependent. If that is the case, then the simple and robust
relations presented here, for the mass of the largest remnant in an
impact event, and for the ejection velocity of the largest remnant, are
appropriate for a new generation of calculations modeling the accretion
of planets in ours and other solar systems.

This work was supported in part by the Swiss National Science Foundation
and by NASA grant NAG5-7245 from the Planetary Geology and Geophysics 
Program.

\section{References}

\pgf Ahrens, T.J., and Rubin, A.M., 1993, Impact-induced tensional 
     failure in rock,
     \emph{J. Geophys. Res.}, \textbf{98}, 1185--1203.

\pgf Asphaug, E. and H. J. Melosh (1993). The Stickney impact of Phobos: a
     dynamical model. {\it Icarus} {\bf 101}, 144--164

\pgf Asphaug, E., 1997,
     Impact origin of the Vesta family,
     \emph{Meteor. and Plan. Sci.}, \textbf{32}, 965--980.

\pgf Asphaug, E., S.J. Ostro, R.S. Hudson, D.J. Scheeres and W. Benz, 1998,
     Disruption of kilometre-sized asteroids by energetic collisions,
     \emph{Nature}, \textbf{393}, 437--440.

\pgf Benz, W., 1989, Smooth Particle Hydrodynamics: a Review, in \emph{
     Numerical Modeling of Nonlinear Stellar Pulsations. Problems and 
     Prospects}, ed. J.R. Buchler, Dordrecht: Kluwer Academic Press, p. 
     269--288.

\pgf Benz, W., and E. Asphaug, 1994,
     Impact simulations with fracture. I. Method and tests,
     \emph{Icarus}, \textbf{107}, 98--116.

\pgf Benz, W., and E. Asphaug, 1995,
     Simulations of brittle solids using smooth particle hydrodynamics
     \emph{Comput. Phys. Comm.}, \textbf{87}, 253--265.

\pgf Davis, D.R., \& Ryan, E.V., 1990, On collisional disruption - Experimental 
     results and scaling laws, {\it Icarus} {\bf 83}, 156.

\pgf Davis, D.R., Ryan, E.V., and P. Farinella, 1997, On how to scale
     disruptive collisions, \emph{Lun. Plan. Science}, \textbf{XXVI},
     319--320.

\pgf Durda, D.D., Greenberg, R., and R. Jedicke, 1998, Collisional models
     and scaling laws: A new interpretation of teh shape of the main-belt
     asteroid size distribution, \emph{Icarus}, \textbf{135}, 431--440.

\pgf Fujiwara, A., Cerroni, P, Davis, D.R., Ryan, E.V., DiMartino, M.,
     Holsapple, K.A., and Housen, K.R., 1989, in \emph{Asteroids II},
     eds. R.P. Binzel, T. Gehrels, M.S. Matthews, University of Arizona
     Press, Tucson.

\pgf Holsapple, K.A., 1994, Catastrophic disruptions and cratering of solar
     system bodies: A review and new results, \emph{Plan. Spac. Science},
     \textbf{42}, 1067--1078.

\pgf Holsapple, K. A., and K. R. Housen, 1986, Scaling laws for the 
     catastrophic collisions of asteroids, \emph{Mem. S.A.It.} {\bf 57}, 
     65-85

\pgf Housen, K.R., and K.A. Holsapple, 1990, On the fragmentation of asteroids
     and planetary satellites, \emph{Icarus}, \textbf{84}, 226--253.

\pgf Housen, K.R., 1991, Laboratory simulations of large-scale fragmentation 
     events, {\it Icarus} {\bf 94}, 180--190.

\pgf Housen, K.R. \& K.A. Holsapple, 1999, Scale effects in strength-dominated
     collisions of rocky asteroids, \emph{Icarus}, this issue.

\pgf Jaeger, J.C., and N.G.W. Cook, 1969,
     Fundamentals of Rock Mechanics, (London:Chapman and Hall).
     
\pgf Lange, M.A. and T.J. Ahrens, 1983, The dynamic tensile strength of ice and 
     ice-silicate mixtures, \emph{J. Geophys. Res} \textbf{88}, 1197--1208.

\pgf Love, S.G., and Ahrens, T.J., 1996, Catastrophic impacts on gravity
     dominated asteroids, \emph{Icarus}, \textbf{124}, 141--155.

\pgf Melosh, H.J., 1989,
     \emph{Impact Cratering: A Geologic Process}, (New York: Oxford University
     Press).
     
\pgf Melosh, H.J., E. Ryan, and E. Asphaug (1992).  Dynamic fragmentation
     in impacts.  {\it J. Geophys. Res.} {\bf 97}: 14,735--14,759.

\pgf Melosh, H.J., and Ryan, E.V., 1997, Asteroids: Shattered not dispersed,
     \emph{Icarus}, \textbf{129}, 562--564.

\pgf Monaghan, J.J., 1992, Smooth Particle Hydrodynamics. \emph{Ann.
     Rev.  Astron. Astrophys.}, \textbf{30}, 543--574.

\pgf Nakamura, A., and A. Fujiwara, 1991,
     Velocity distribution of fragments formed in a simulated collisional
     disruption, \emph{Icarus}, \textbf{92}, 132--146.
     
\pgf O'Keefe, J.D. and T.J. Ahrens, 1982a, The interaction of the 
     Cretaceous/Tertiary extinction bolide with the atmosphere, ocean 
     and solid earth. {\it Geol. Soc. Amer.} Special Papers {\bf 190}, 
     103--120.
     
\pgf O'Keefe, J.D. and T.J. Ahrens, 1982b, Cometary and meteorite 
     swarm impact on planetary surfaces, {\it J. Geophys. Res.} {\bf 87}, 
     6668--6680.
     
\pgf Ryan, E.V., Hartmann, W.K., \& Davis, D.R., Impact 
     experiments III - Catastrophic fragmentation of aggregate targets 
     and relation to asteroids, 1991, {\it Icarus} {\bf 94}, 283--298

\pgf Ryan, E.V., and Melosh, H.J., 1998, Impact fragmentation: From the
     laboratory to asteroids, Icarus, 133, 1--24
     
\pgf Shoemaker, E. M., 1962, Interpenetration of lunar craters.  In 
     Z. Kopal (Ed.), \emph{Physics and Astronomy of the Moon}, Academic 
     Press, New York and London, pp. 283--359.     

\pgf Tillotson, J. H., 1962, Metallic equations of state for hypervelocity 
     impact, \emph{Rep. GA-3216}, July 18, Gen. At., San Diego, California

\pgf Weibull, W. A., 1939,
     A statistical theory of the strength of materials (transl.),
     \emph{Ingvetensk. Akad. Handl.}, \textbf{151} (Stockholm), 5--45.

\newpage

\begin{table}[h!]
   \begin{center}
   \caption{Material-dependent constants}
   \label{tab:matcon}
   \footnotesize
   \begin{tabular}{llllll}
   \hline\hline
       & $\mu$  & Y      & $E_{melt}$  & k         & m  \\
       & erg/cc & erg/g  & erg/g       & cm$^{-3}$ &    \\
   \hline
basalt & $2.27 \; 10^{11}$ & $3.5 \; 10^{10}$ & $3.4 \; 10^{10}$  & $4.0 \;
10^{29} \ ^a$  & 9.0 \ $^a$ \\
ice    & $2.80 \; 10^{10}$ & $1.0 \; 10^{10}$ & $7.0 \; 10^{9}$  & $1.4 \;
10^{32} \ ^b$ & 9.6 \ $^b$ \\
\hline\hline \\
\multicolumn{6}{l}{$^a$ From calibrating numerical simulations to laboratory} \\
\multicolumn{6}{l}{\ \ \ experiments; Benz \& Asphaug (1995)}\\
\multicolumn{6}{l}{$^b$ Our fit to stress-strain rate data published in Lange} \\
\multicolumn{6}{l}{\ \ \  \& Ahrens (1983)}\\
\end{tabular}
\normalsize
\end{center}
\end{table}

\newpage

\begin{table}[h!]
   \begin{center}
   \caption{Tillotson eos parameters}
   \label{tab:tillotson}
   \scriptsize
   \begin{tabular}{lllllllllll}
   \hline\hline
       & $\rho_0$ & A          & B          & $E_{0}$ & E$_{\rm iv}$ & E$_{\rm
cv}$ &
         a & b & $\alpha$ & $\beta$ \\
       & g/cc  & erg/cc & erg/cc & erg/g   & erg/g    & erg/g    &   &   &
                   &         \\
   \hline
basalt & $2.7$  & $2.67 \; 10^{11}$ & $2.67 \; 10^{11}$ & $4.87 \; 10^{12}$ &
$4.72 \; 10^{10}$ &
         $1.82 \; 10^{11}$ & $0.5$ & $1.50$ & $5.0$ & $5.0$ \ $^a$ \\
ice    & $0.917$& $9.47 \; 10^{10}$ & $9.47 \; 10^{10}$ & $1.00 \; 10^{11}$ &
$7.73 \; 10^9$    &
         $3.04 \; 10^{10}$ & $0.3$ & $0.1$ & $10.0$ & $5.0$ \ $^b$ \\
\hline\hline \\
\multicolumn{11}{l}{$^a$ Tillotson parameters as published for lunar gabbroic
                        anorthosite (O'Keefe \& Ahrens 1982a), substituting}\\
\multicolumn{11}{l}{\ \ \ the basalt reference density and bulk modulus as reported
                        by Nakamura \& Fujiwara (1991), from whose} \\
\multicolumn{11}{l}{\ \ \ work our basalt fracture coefficients (Table 1) are derived.}\\
\multicolumn{11}{l}{$^b$ Tillotson parameters as expressed for water ice (O'Keefe
                         \& Ahrens 1982b), again substituting the ice}\\
\multicolumn{11}{l}{\ \ \ reference density and bulk modulus as reported by Lange 
                          \& Ahrens (1983), from whose work our ice}\\ 
\multicolumn{11}{l}{\ \ \ fracture coefficients are derived.}\\
\end{tabular}
\normalsize
\end{center}
\end{table}

\newpage

\begin{table}[h!]
   \begin{center}
   \caption{Fit constants for $\bar{Q}^*_D$}
   \label{tab:qbarcon}
   \footnotesize
   \begin{tabular}{cccccc}
   \hline\hline
    material & $v_{\rm impact}$  & $Q_0$        &   B               & a     &
b     \\
             &    km/s           & erg/g        & erg\,cm$^3$/g$^2$ &       &
     \\
   \hline
   basalt    &    5              & 9.0\,10$^7$ &   0.5             & -0.36 &
1.36  \\
   basalt    &    3              & 3.5\,10$^7$ &   0.3             & -0.38 &
1.36  \\
   ice       &    3              & 1.6\,10$^7$ &   1.2             & -0.39 &
1.26  \\
   ice       &    0.5            & 7.0\,10$^7$ &   2.1             & -0.45 &
1.19  \\
\hline\hline
\end{tabular}
\normalsize
\end{center}
\end{table}

\newpage

\begin{table}[ht!]
   \begin{center}
   \caption{Radius of the weakest object}
   \label{tab:rweak}
   \footnotesize
   \begin{tabular}{ccc}
   \hline\hline
    material & $v_{\rm impact}$  & $R_{\rm weak}$ \\
             &    km/s           & m              \\
   \hline
   basalt    &    5              & 163            \\
   basalt    &    3              & 117            \\
   ice       &    3              & 102            \\
   ice       &    0.5            & 213            \\
\hline\hline
\end{tabular}
\normalsize
\end{center}
\end{table}

\newpage

\section*{Figure Captions}

\begin{figure}[htb!]
  \begin{flushleft}
  \caption{In the strength dominated regime, the largest remnant is a
           single intact fragment while in the gravity dominated regime
           the largest remnant is a gravitationally bound aggregate of
           fragments of various sizes.}
  \label{fig:impcart}
  \end{flushleft}
\end{figure}
\begin{figure}[htb!]
  \begin{flushleft}
  \caption{Catastrophic disruption thresholds for a basalt target
           and 3 km/s impact velocity. Each set of connected dots
           represent one projectile angle of incidence starting with
           0 (bottom curve), 30, 45, 60, and 75 (top curve) degrees.}
  \label{fig:qbas3kms}
  \end{flushleft}
\end{figure}
\begin{figure}[htb!]
  \begin{flushleft}
  \caption{Catastrophic disruption thresholds for a basalt target
           and 5 km/s impact velocity. Each set of connected dots
           represent one projectile angle of incidence starting with
           0 (bottom curve), 30, 45, 60, and 75 (top curve) degrees.}
  \label{fig:qbas5kms}
  \end{flushleft}
\end{figure}
\begin{figure}[htb!]
  \begin{flushleft}
  \caption{Catastrophic disruption thresholds for an icy target
           and 3 km/s impact velocity. See Fig(\ref{fig:qbas3kms})
           for caption}
  \label{fig:qice3kms}
  \end{flushleft}
\end{figure}
\begin{figure}[htb!]
  \begin{flushleft}
  \caption{Catastrophic disruption thresholds for an icy target
           and 0.5 km/s impact velocity. See Fig(\ref{fig:qbas3kms})
           for caption}
  \label{fig:qice05kms}
  \end{flushleft}
\end{figure}
\begin{figure}[htb!]
  \begin{flushleft}
  \caption{Mean catastrophic disruption threshold $\bar{Q}^*_D$
           for a random distribution of impact parameters (see text
           for details) in the case of basalt targets}
  \label{fig:qbasmean}
  \end{flushleft}
\end{figure}
\begin{figure}[htb!]
  \begin{flushleft}
  \caption{Mean catastrophic disruption threshold $\bar{Q}^*_D$
           for a random distribution of impact parameters (see text
           for details) in the case of icy targets}
  \label{fig:qicemean}
  \end{flushleft}
\end{figure}
\begin{figure}[htb!]
  \begin{flushleft}
  \caption{Comparison between the mean catastrophic disruption threshold
           $\bar{Q}^*_D$ (basalt targets, $v=3$km/s) determined in this
           work (heavy line) and other determinations in the literature}
  \label{fig:compq}
  \end{flushleft}
\end{figure}
\begin{figure}[htb!]
  \begin{flushleft}
  \caption{Mass of largest remnant (in terms of original target mass)
           as a function of impact energy per gram of target material
           normalized to \qcrit for icy targets and collisions occurring
           at a) 0.5 km/s b) 3km/s. Different symbols correspond to
           different target sizes}
  \label{fig:icemfq}
  \end{flushleft}
\end{figure}
\begin{figure}[htb!]
  \begin{flushleft}
  \caption{Mass of largest remnant (in terms of original target mass)
           as a function of impact energy per gram of target material
           normalized to \qcrit for basaltic targets and collisions
           occurring at a) 3 km/s b) 5 km/s. Different symbols correspond
           to different target sizes}
  \label{fig:basmfq}
  \end{flushleft}
\end{figure}
\begin{figure}[htb!]
  \begin{flushleft}
  \caption{The mass of the largest intact fragment (normalized to mass
           of the parent body) when the threshold critical disruption occurs
           (\qcrit) for basalt targets. When gravity goes to zero, $M_{lif}$
           goes to 0.5 by definition.  The results for different angles of
           incidence are shown by using different symbols}
  \label{fig:fraginbas}
  \end{flushleft}
\end{figure}
\begin{figure}[htb!]
  \begin{flushleft}
  \caption{The mass of the largest intact fragment (normalized to mass
           of the parent body) when the threshold critical disruption occurs
           (\qcrit) for icy targets. When gravity goes to zero, $M_{lif}$
           goes to 0.5 by definition.  The results for different angles of
           incidence are shown by using different symbols}
  \label{fig:fraginice}
  \end{flushleft}
\end{figure}
\begin{figure}[htb!]
  \begin{flushleft}
  \caption{Ejection velocities of largest remnant in 5 km/s collisions
           involving basalt targets. Different symbols correspond to 
           different parent body sizes. In the lower panel, the velocities 
           have been normalized to the parent body's escape velocity and
           only the results for initial bodies with $R_{pb} \ge 1$km 
           are shown}
  \label{fig:vbas5kms}
  \end{flushleft}
\end{figure}
\begin{figure}[htb!]
  \begin{flushleft}
  \caption{Ejection velocities of largest remnant in 3 km/s collisions
           involving basalt targets. Different symbols correspond to 
           different parent body sizes. In the lower panel, the velocities 
           have been normalized to the parent body's escape velocity and
           only the results for initial bodies with $R_{pb} \ge 1$km 
           are shown}
  \label{fig:vbas3kms}
  \end{flushleft}
\end{figure}
\begin{figure}[htb!]
  \begin{flushleft}
  \caption{Ejection velocities of largest remnant in 3 km/s collisions
           involving icy targets. Different symbols correspond to 
           different parent body sizes. In the lower panel, the velocities 
           have been normalized to the parent body's escape velocity and
           only the results for initial bodies with $R_{pb} \ge 1$km 
           are shown}
  \label{fig:vice3kms}
  \end{flushleft}
\end{figure}
\begin{figure}[htb!]
  \begin{flushleft}
  \caption{Ejection velocities of largest remnant in 0.5 km/s collisions
           involving icy targets. Different symbols correspond to 
           different parent body sizes. In the lower panel, the velocities 
           have been normalized to the parent body's escape velocity and
           only the results for initial bodies with $R_{pb} \ge 1$km 
           are shown}
  \label{fig:vice05kms}
  \end{flushleft}
\end{figure}
\begin{figure}[htb!]
  \begin{flushleft}
  \caption{Ejection velocities of largest remnant as a function of
           impactor radius normalized to target radius in collisions
           involving basalt targets.}
  \label{fig:basvr}
  \end{flushleft}
\end{figure}
\begin{figure}[htb!]
  \begin{flushleft}
  \caption{Ejection velocities of largest remnant as a function of
           impactor radius normalized to target radius in collisions
           involving icy targets.}
  \label{fig:icevr}
  \end{flushleft}
\end{figure}

\clearpage
\newpage
\setcounter{figure}{0}

%
%
\begin{figure}[p!]
  \begin{flushleft}
  \epsfig{file=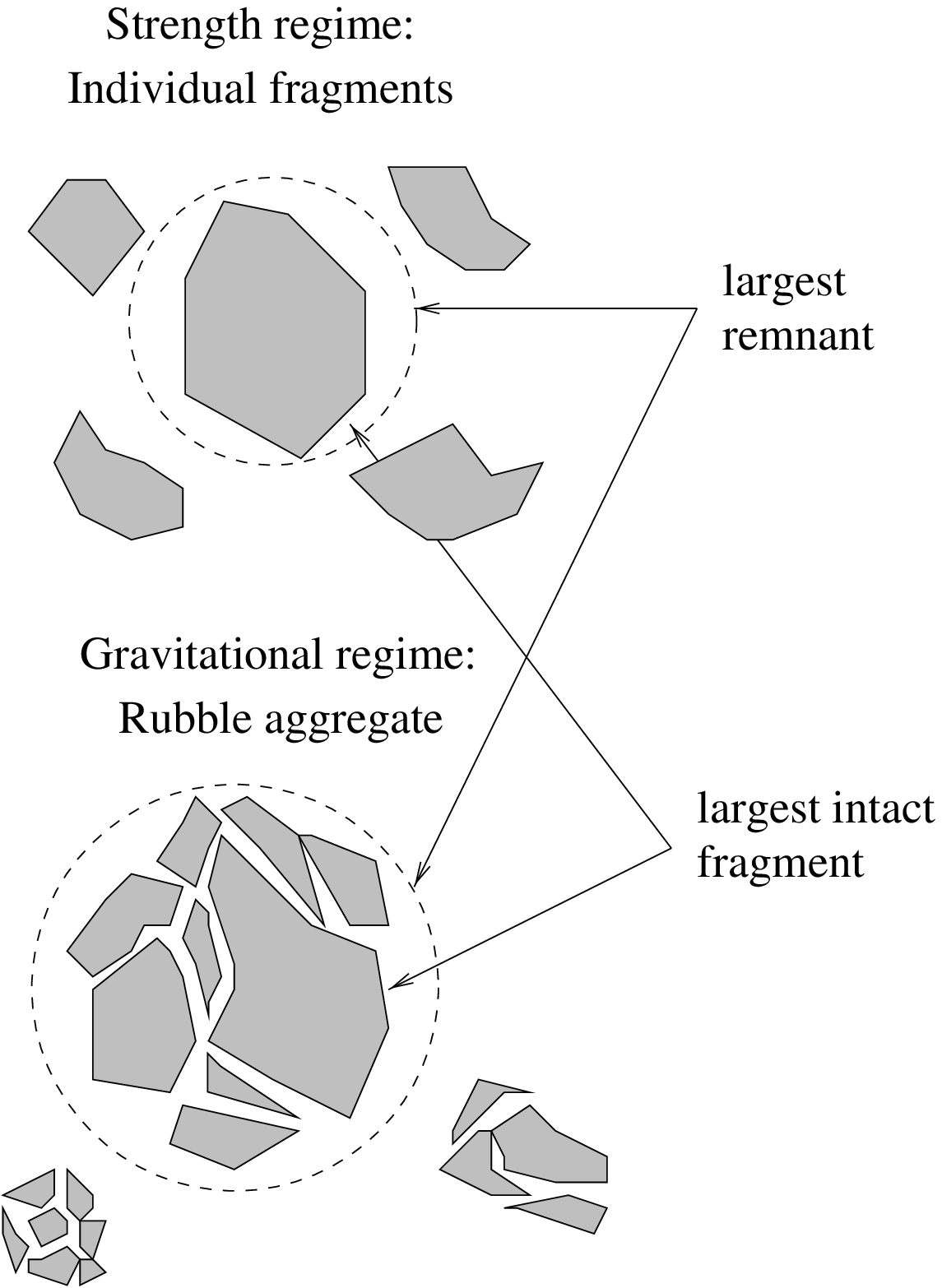,width=12cm}
  \caption{\ }
  \end{flushleft}
\end{figure}
\begin{figure}[p!]
  \begin{flushleft}
  \epsfig{file=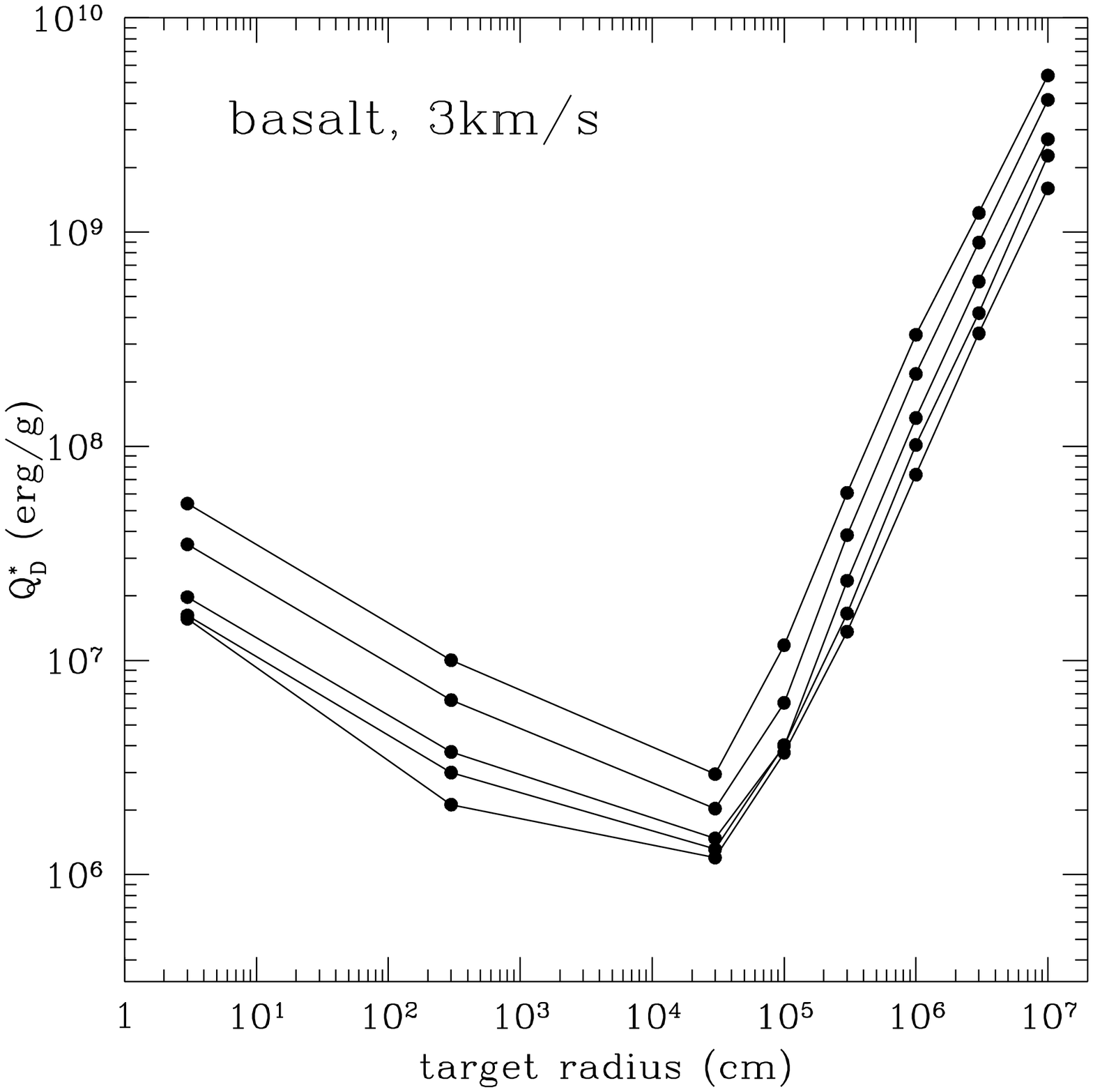,width=12cm}
  \caption{\ }
  \end{flushleft}
\end{figure}
\begin{figure}[p!]
  \begin{flushleft}
  \epsfig{file=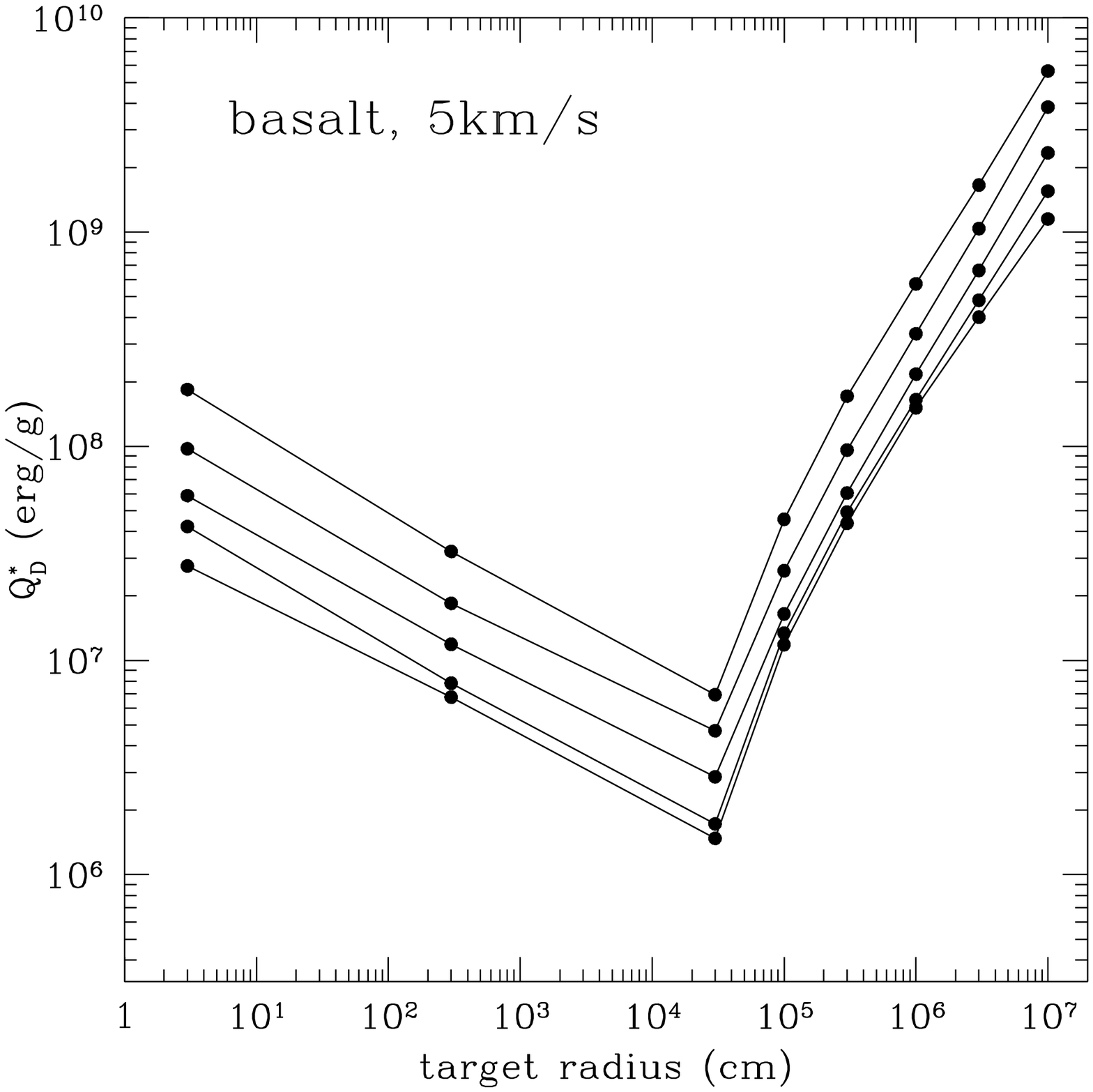,width=12cm}
  \caption{\ }
  \end{flushleft}
\end{figure}
\begin{figure}[p!]
  \begin{flushleft}
  \epsfig{file=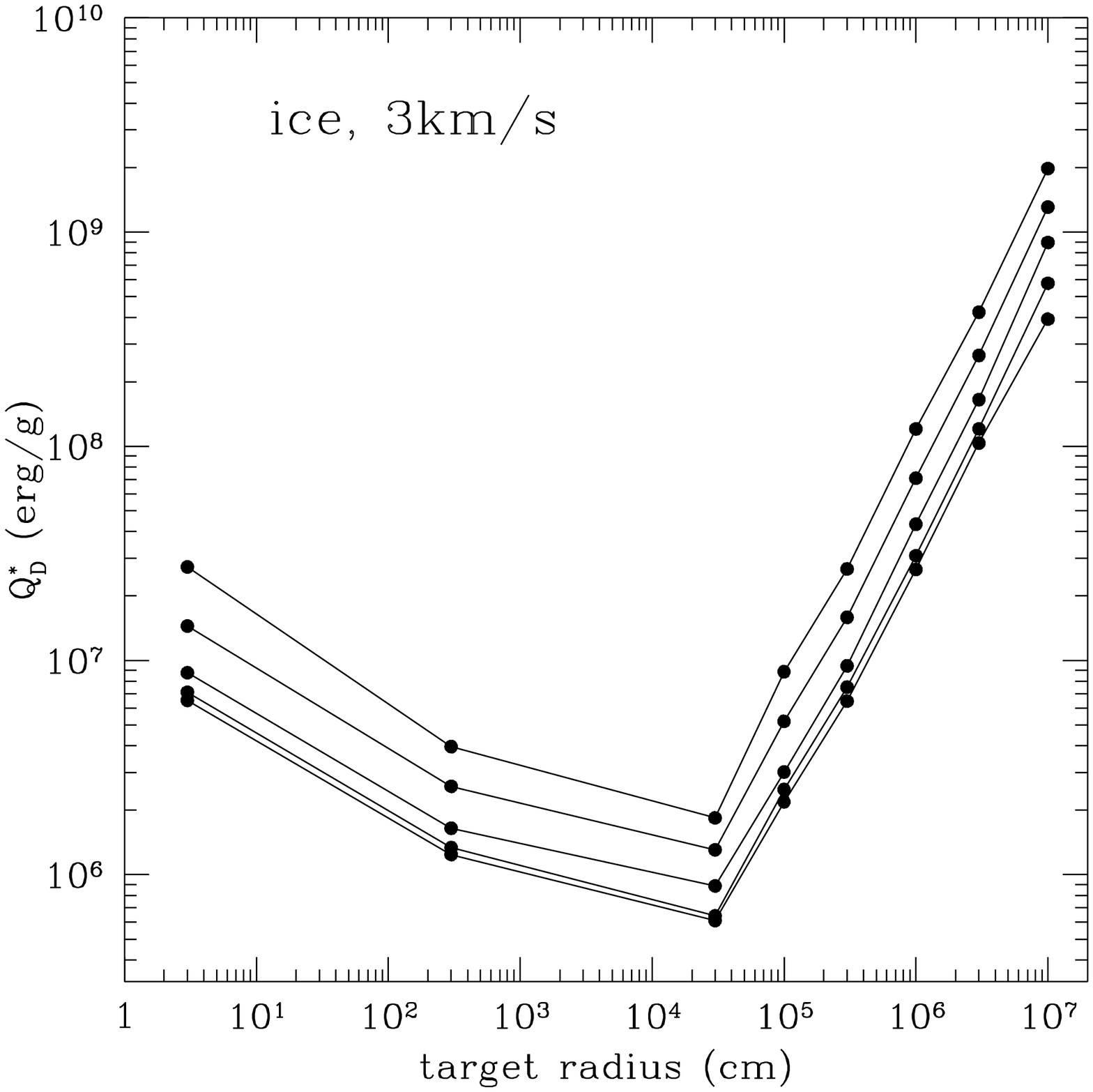,width=12cm}
  \caption{\ }
  \end{flushleft}
\end{figure}
\begin{figure}[p!]
  \begin{flushleft}
  \epsfig{file=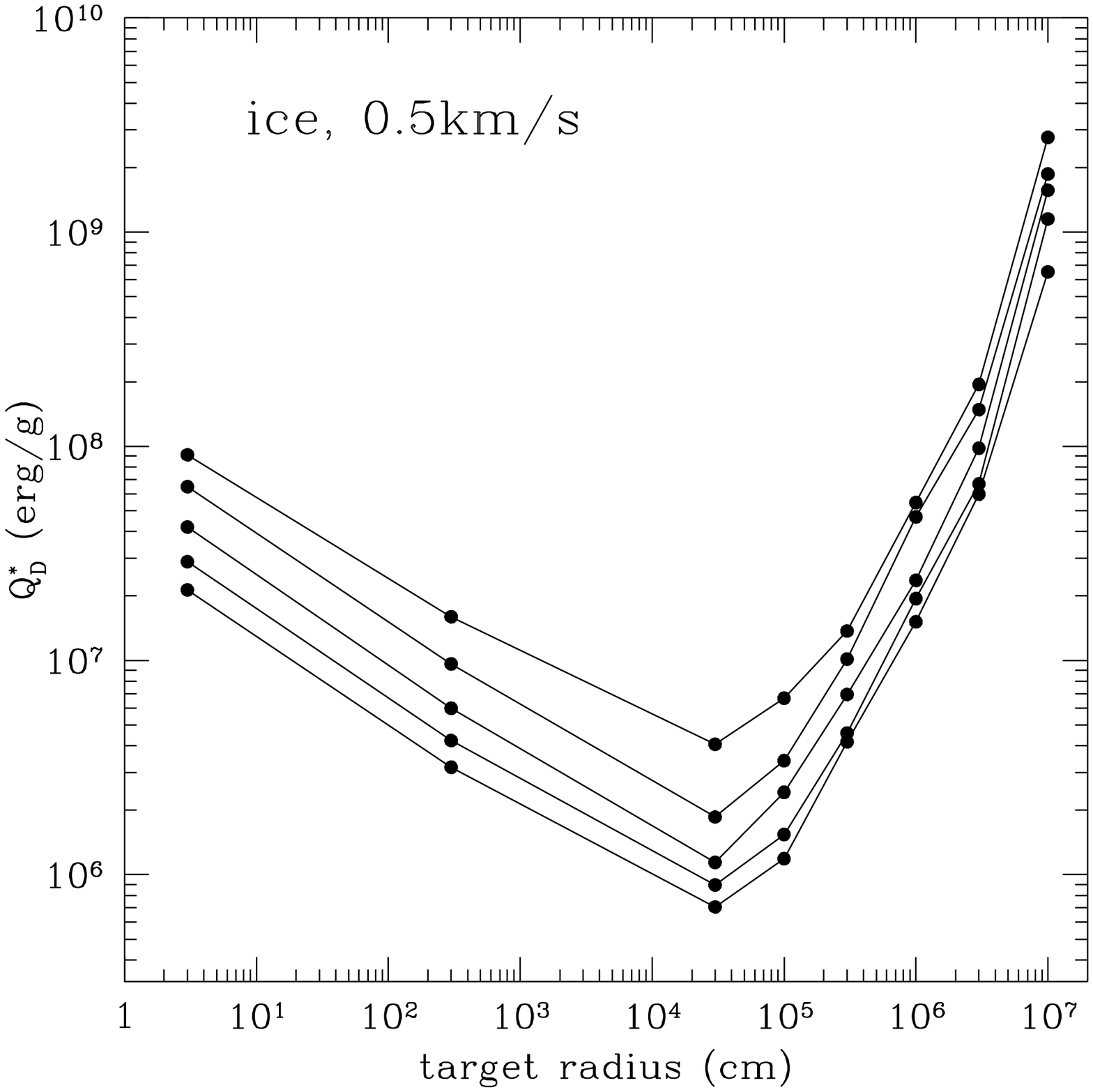,width=12cm}
  \caption{\ }
  \end{flushleft}
\end{figure}
\begin{figure}[p!]
  \begin{flushleft}
  \epsfig{file=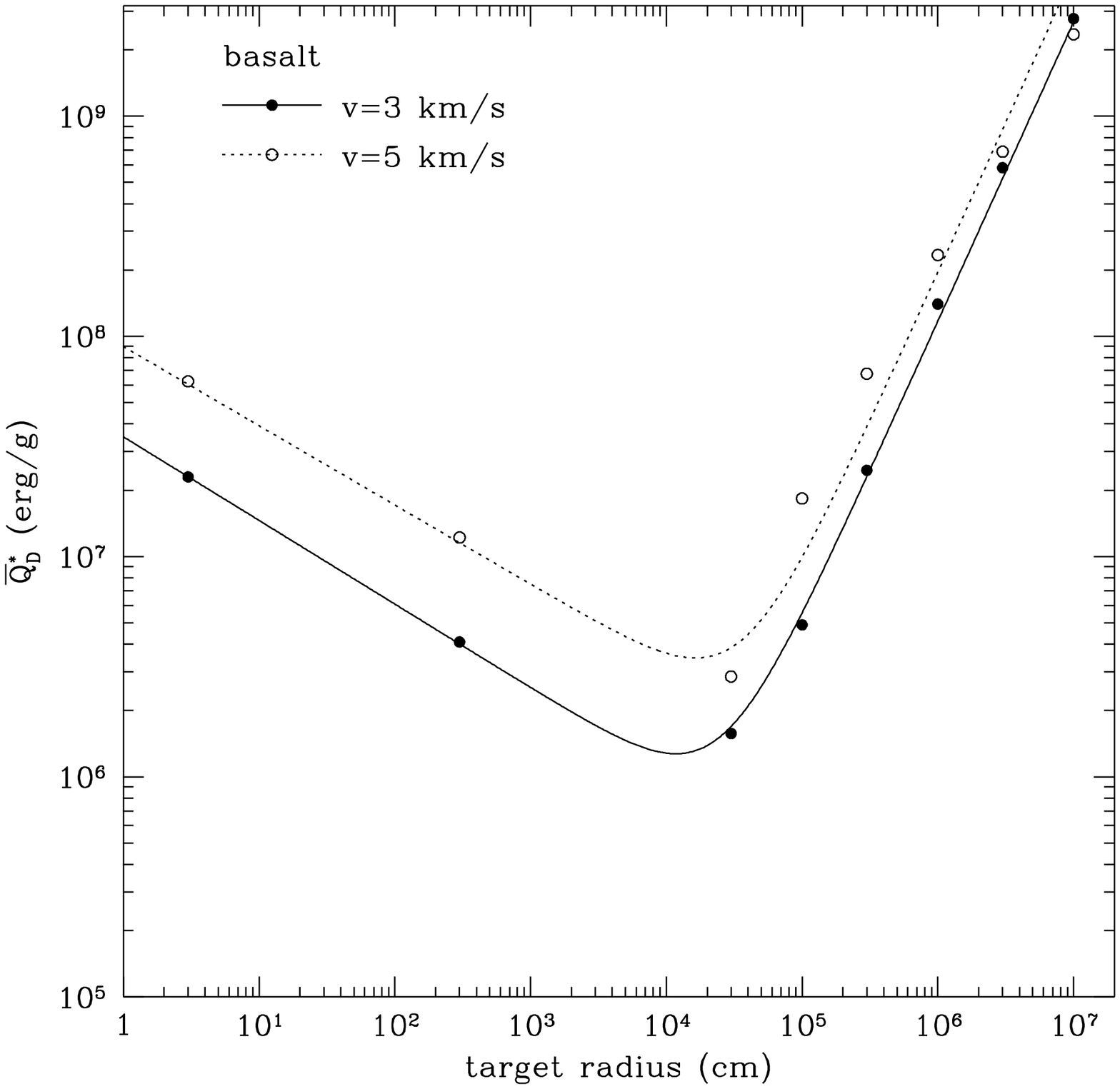,width=12cm}
  \caption{\ }
  \end{flushleft}
\end{figure}
\begin{figure}[p!]
  \begin{flushleft}
  \epsfig{file=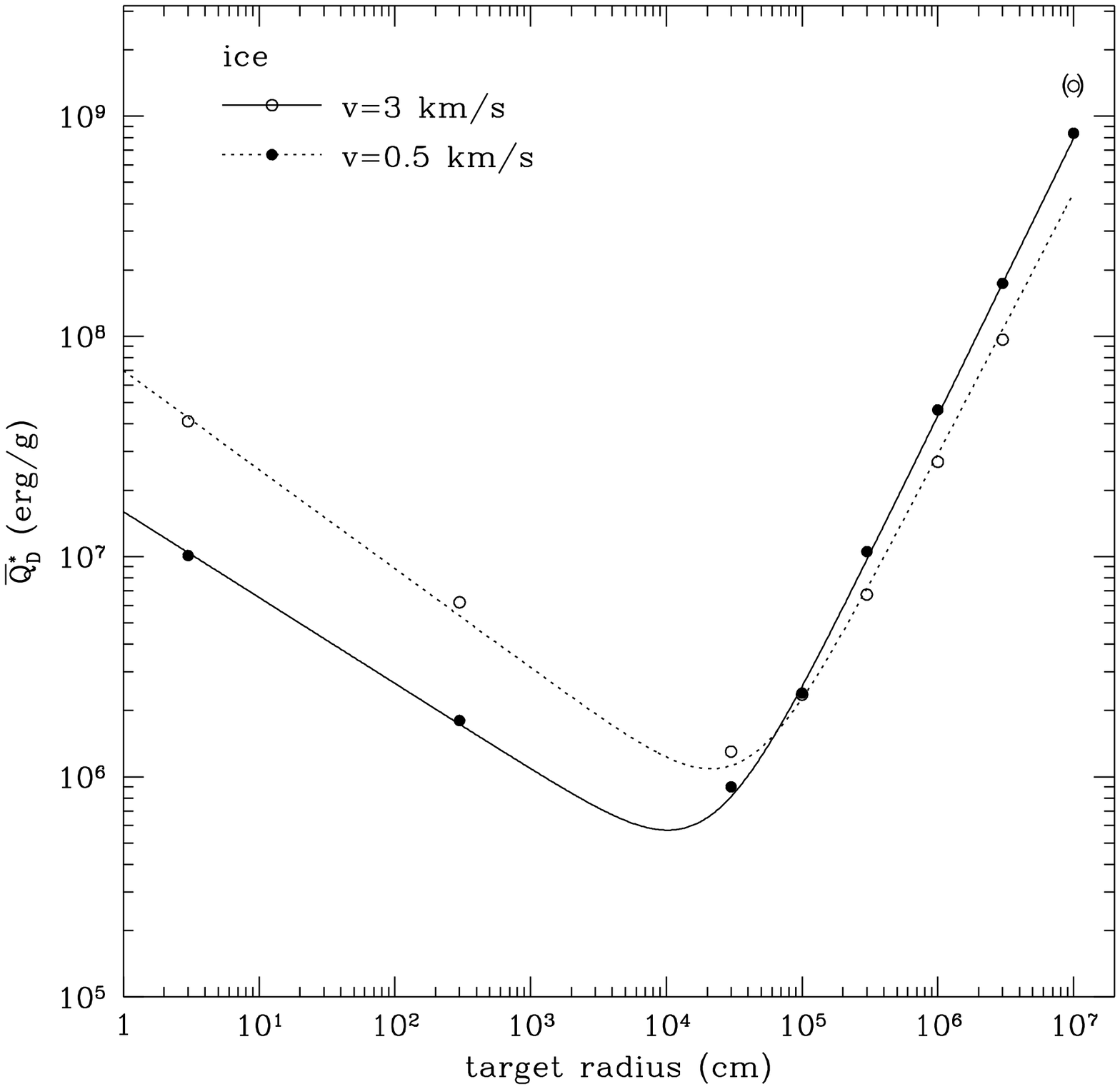,width=12cm}
  \caption{\ }
  \end{flushleft}
\end{figure}
\begin{figure}[p!]
  \begin{flushleft}
  \epsfig{file=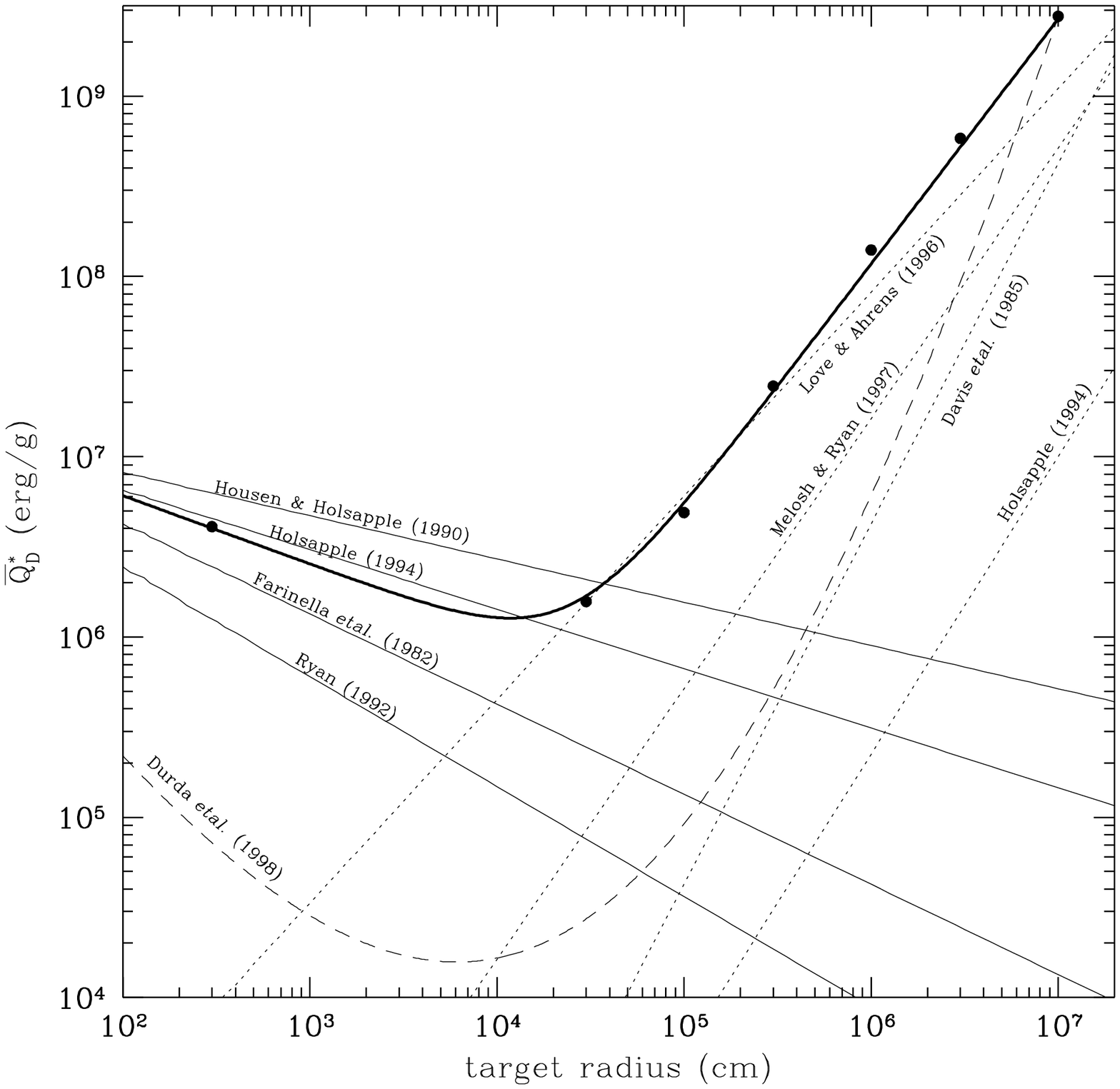,width=12cm}
  \caption{\ }
  \end{flushleft}
\end{figure}
\begin{figure}[p!]
  \begin{flushleft}
  \epsfig{file=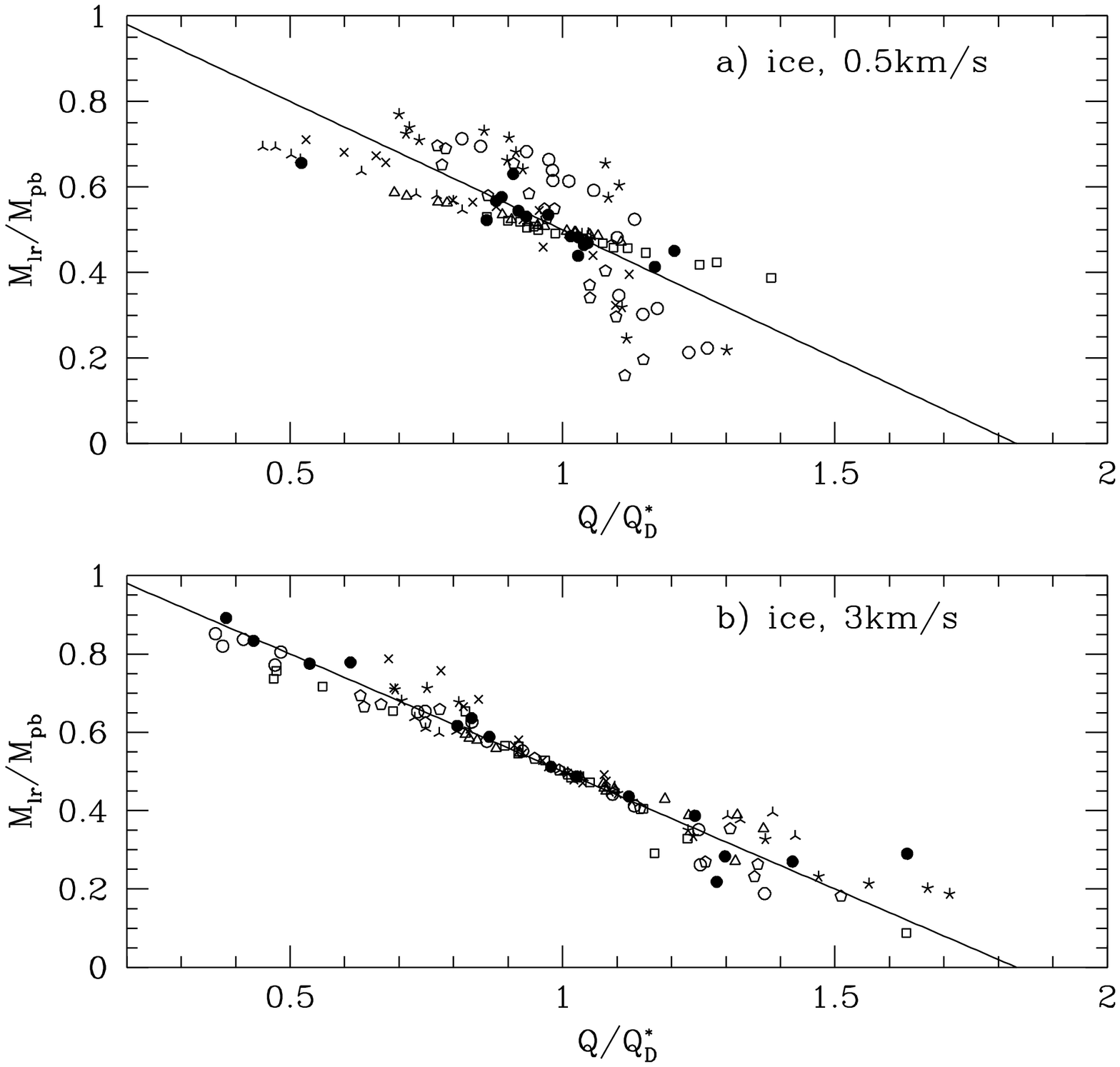,width=12cm}
  \caption{\ }
  \end{flushleft}
\end{figure}
\begin{figure}[p!]
  \begin{flushleft}
  \epsfig{file=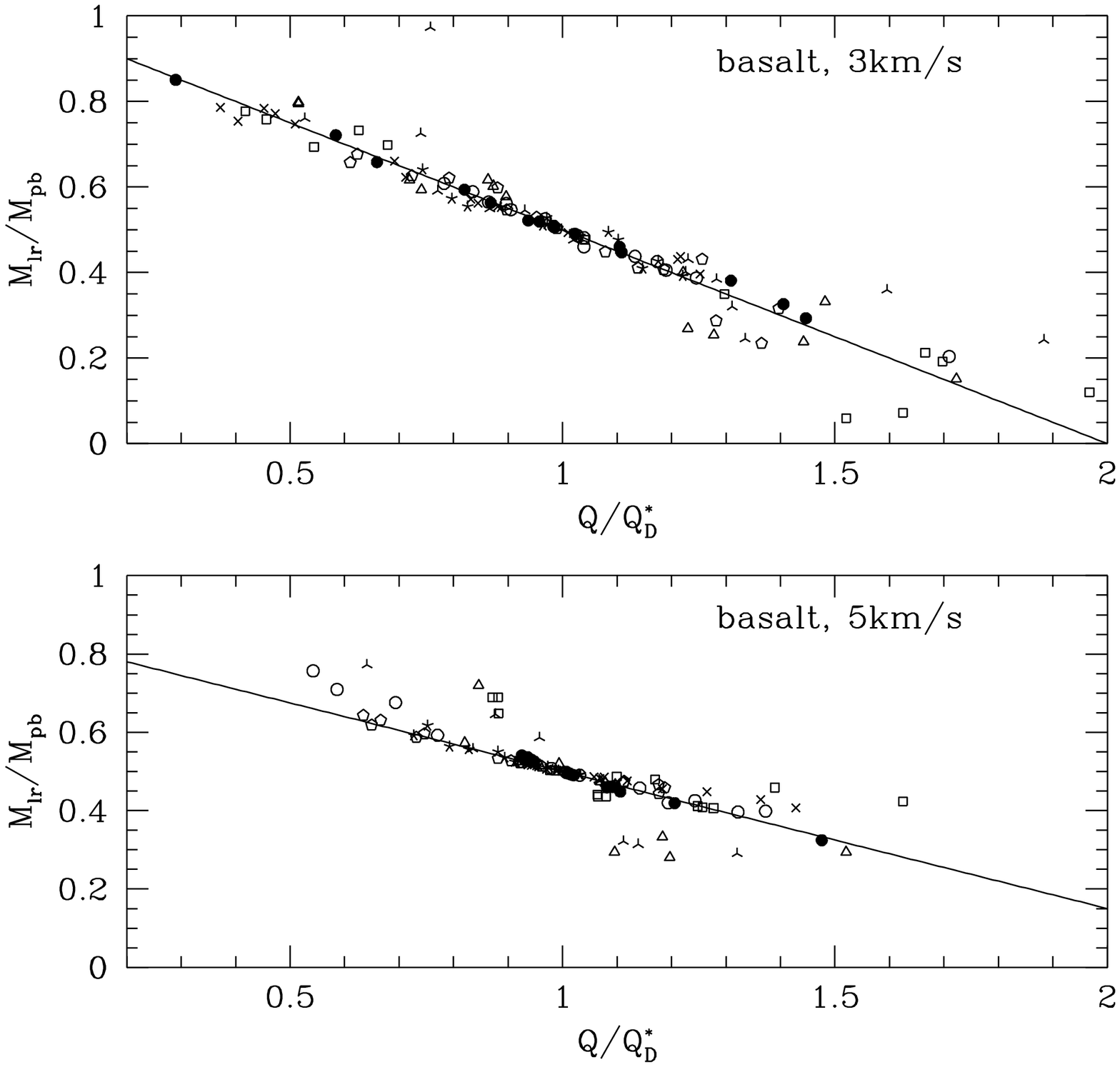,width=12cm}
  \caption{\ }
  \end{flushleft}
\end{figure}
\begin{figure}[p!]
  \begin{flushleft}
  \epsfig{file=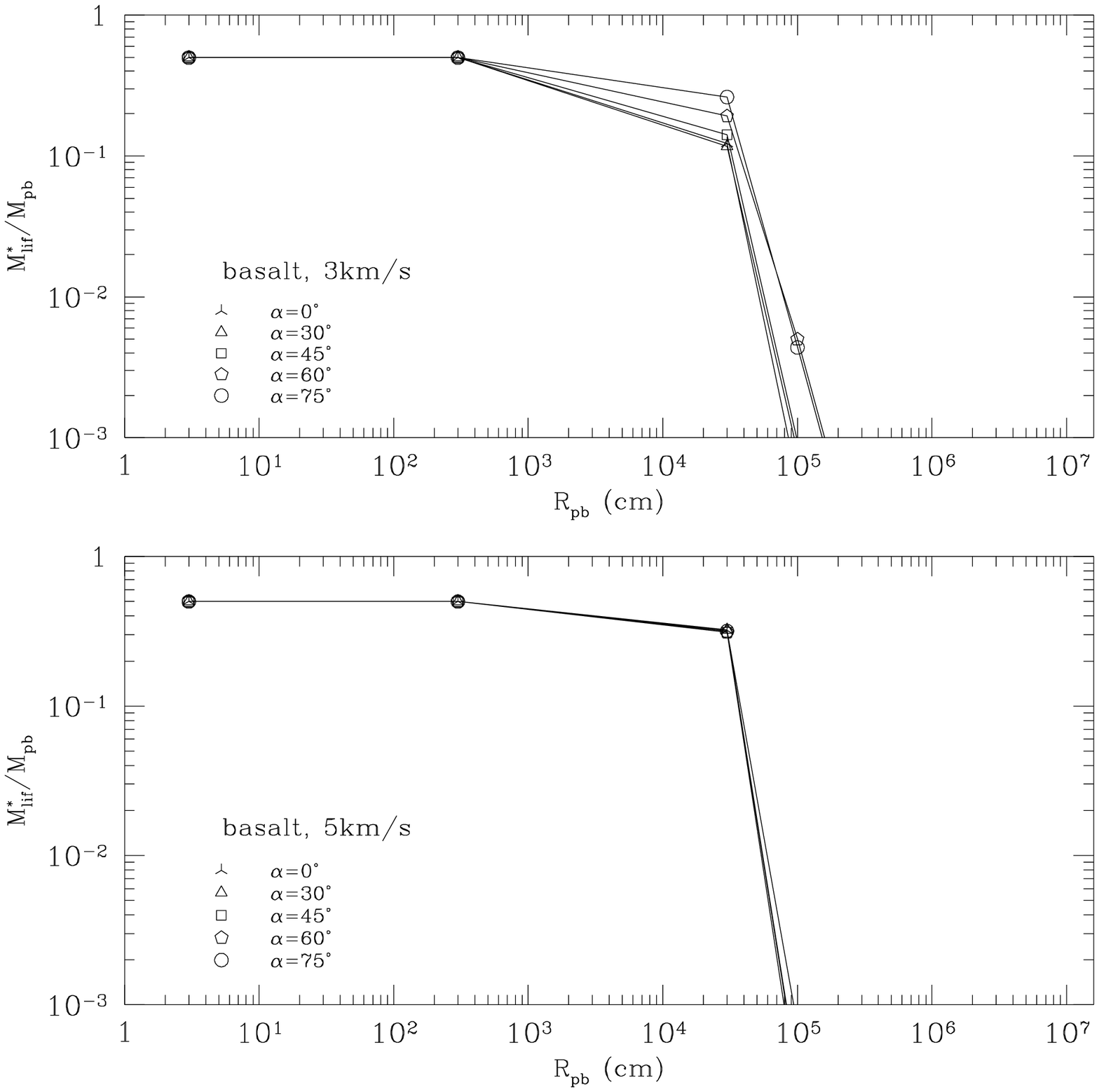,width=12cm}
  \caption{\ }
  \end{flushleft}
\end{figure}
\begin{figure}[p!]
  \begin{flushleft}
  \epsfig{file=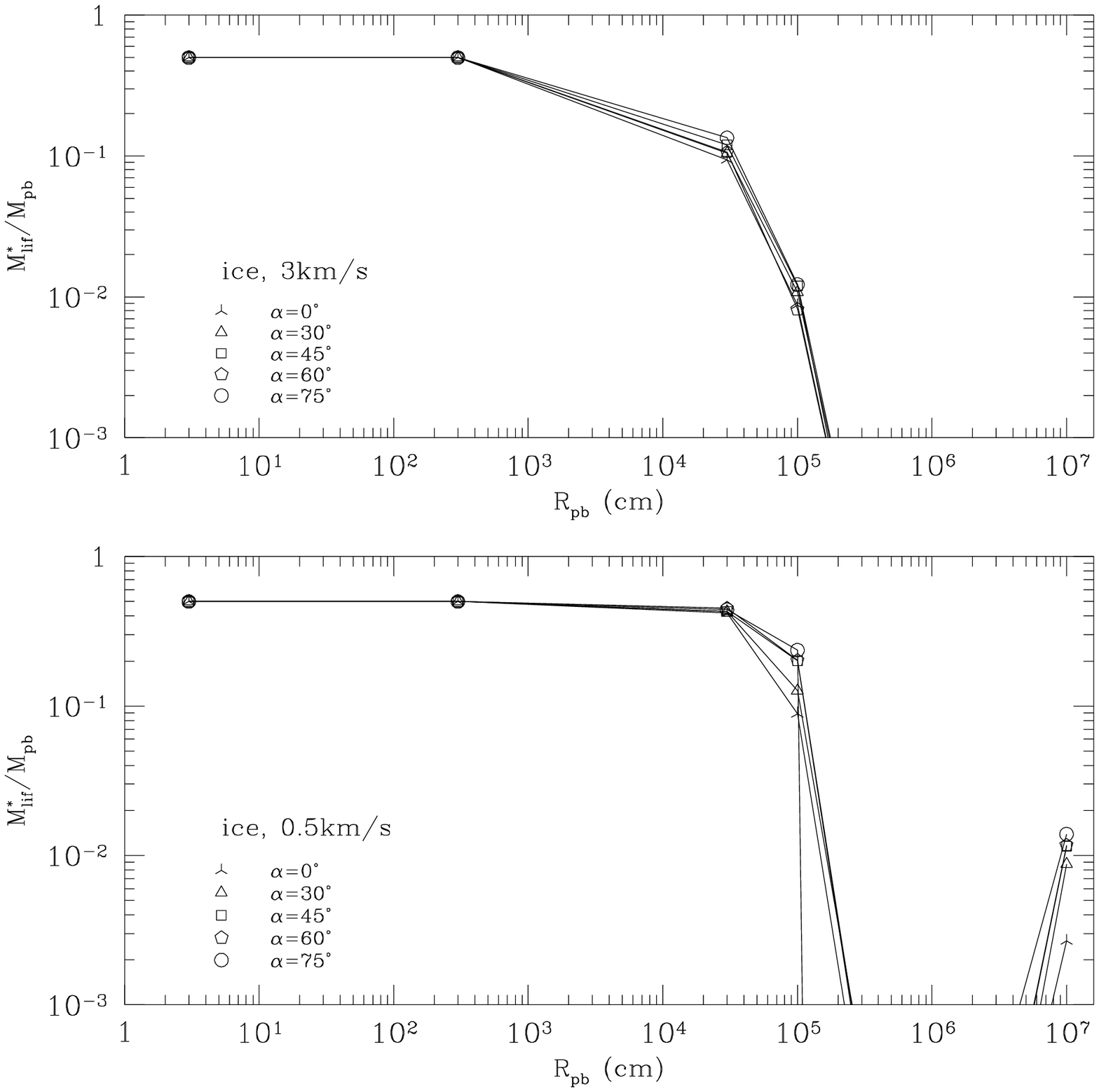,width=12cm}
  \caption{\ }
  \end{flushleft}
\end{figure}
\begin{figure}[p!]
  \begin{flushleft}
  \epsfig{file=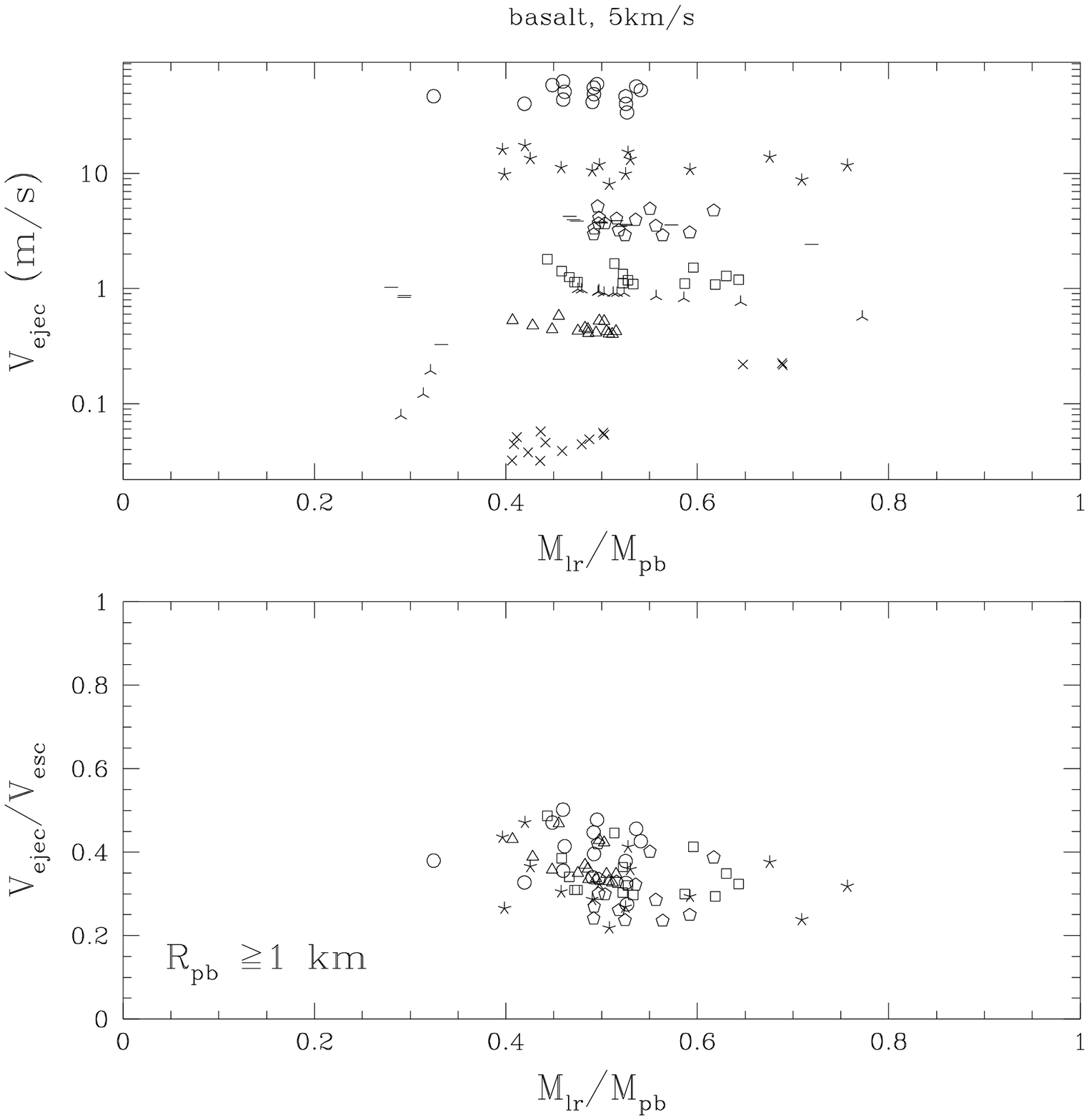,width=12cm}
  \caption{\ }
  \end{flushleft}
\end{figure}
\begin{figure}[p!]
  \begin{flushleft}
  \epsfig{file=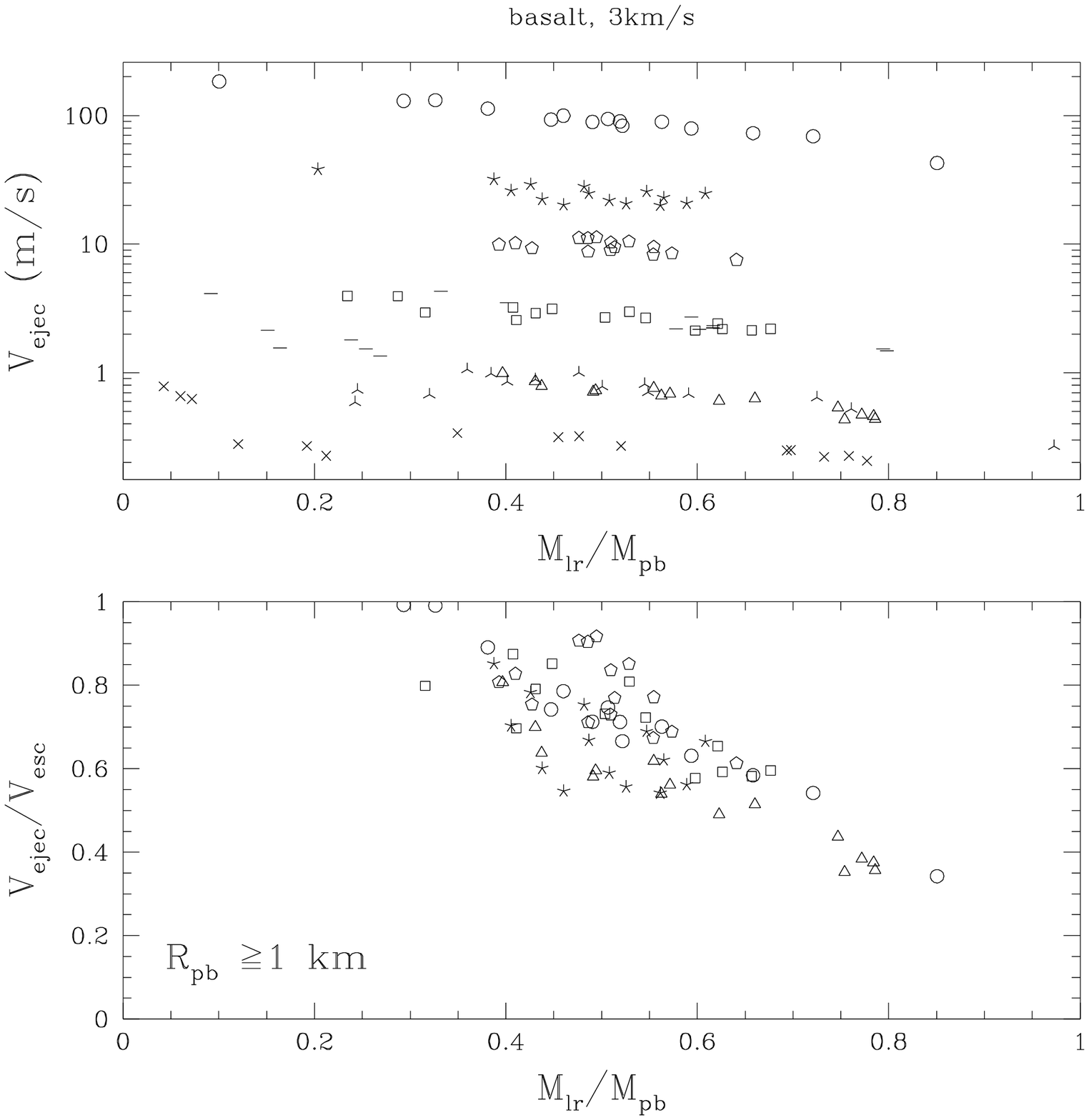,width=12cm}
  \caption{\ }
  \end{flushleft}
\end{figure}
\begin{figure}[p!]
  \begin{flushleft}
  \epsfig{file=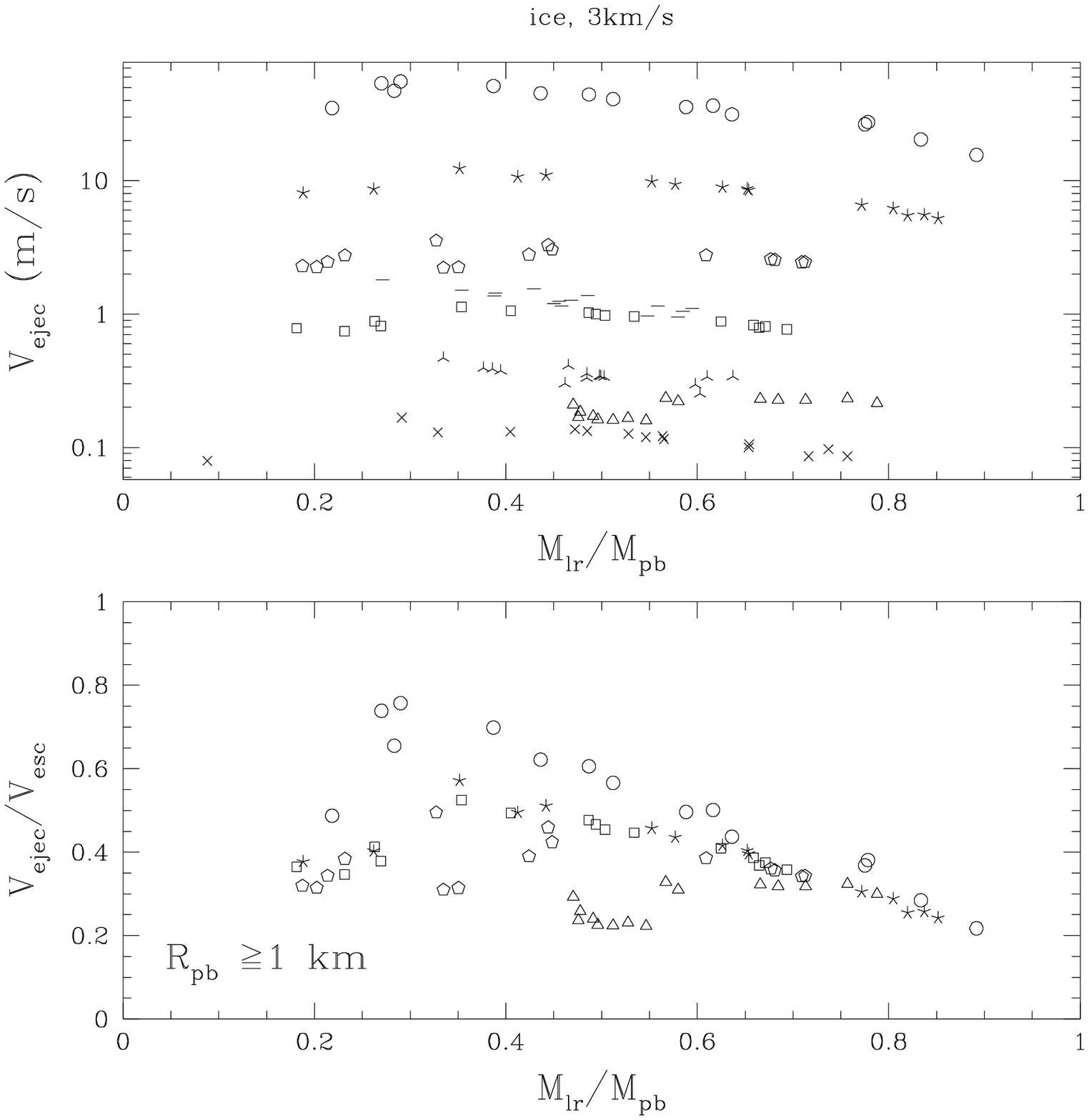,width=12cm}
  \caption{\ }
  \end{flushleft}
\end{figure}
\begin{figure}[p!]
  \begin{flushleft}
  \epsfig{file=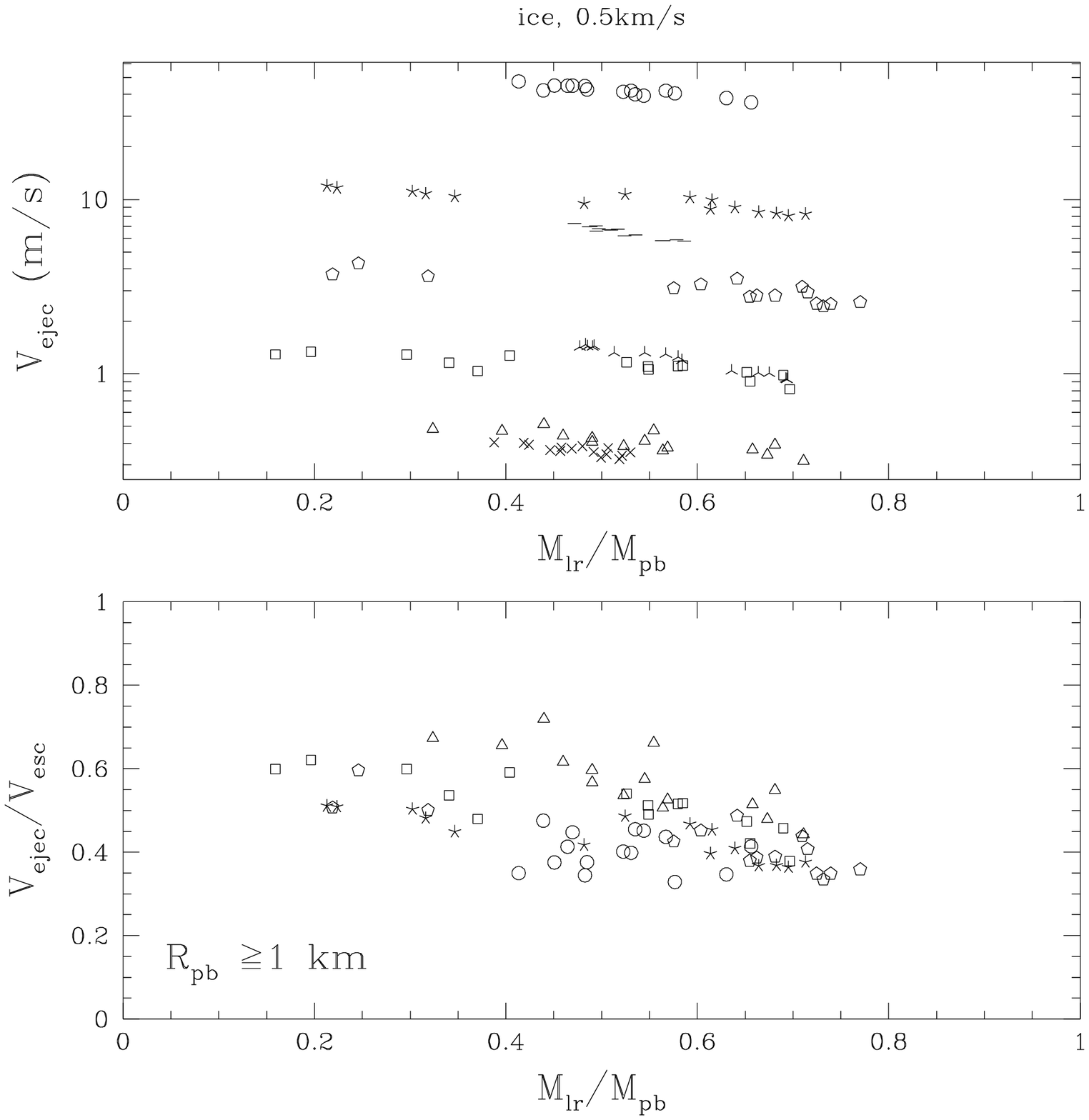,width=12cm}
  \caption{\ }
  \end{flushleft}
\end{figure}
\begin{figure}[p!]
  \begin{flushleft}
  \epsfig{file=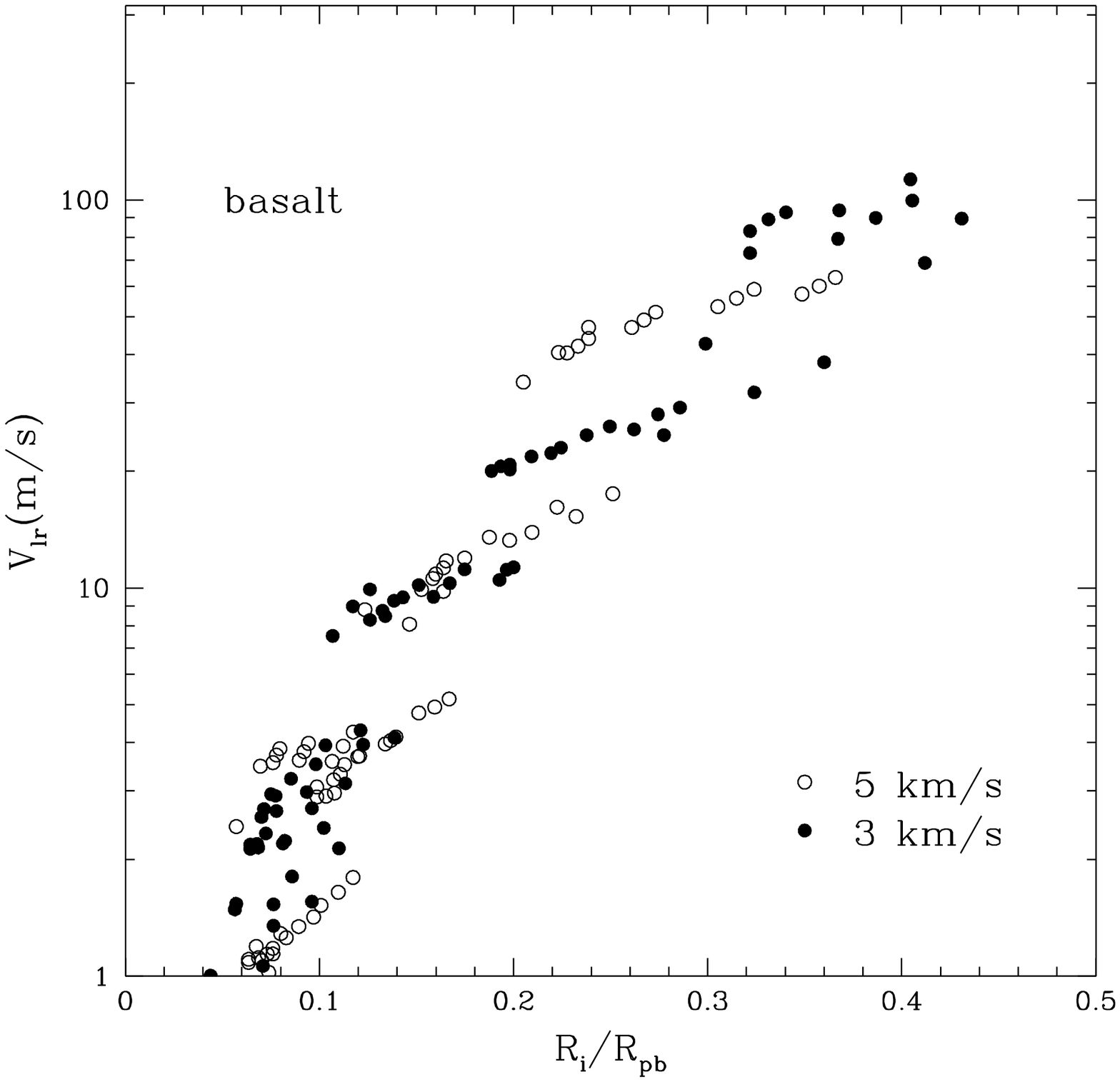,width=12cm}
  \caption{\ }
  \end{flushleft}
\end{figure}
\begin{figure}[p!]
  \begin{flushleft}
  \epsfig{file=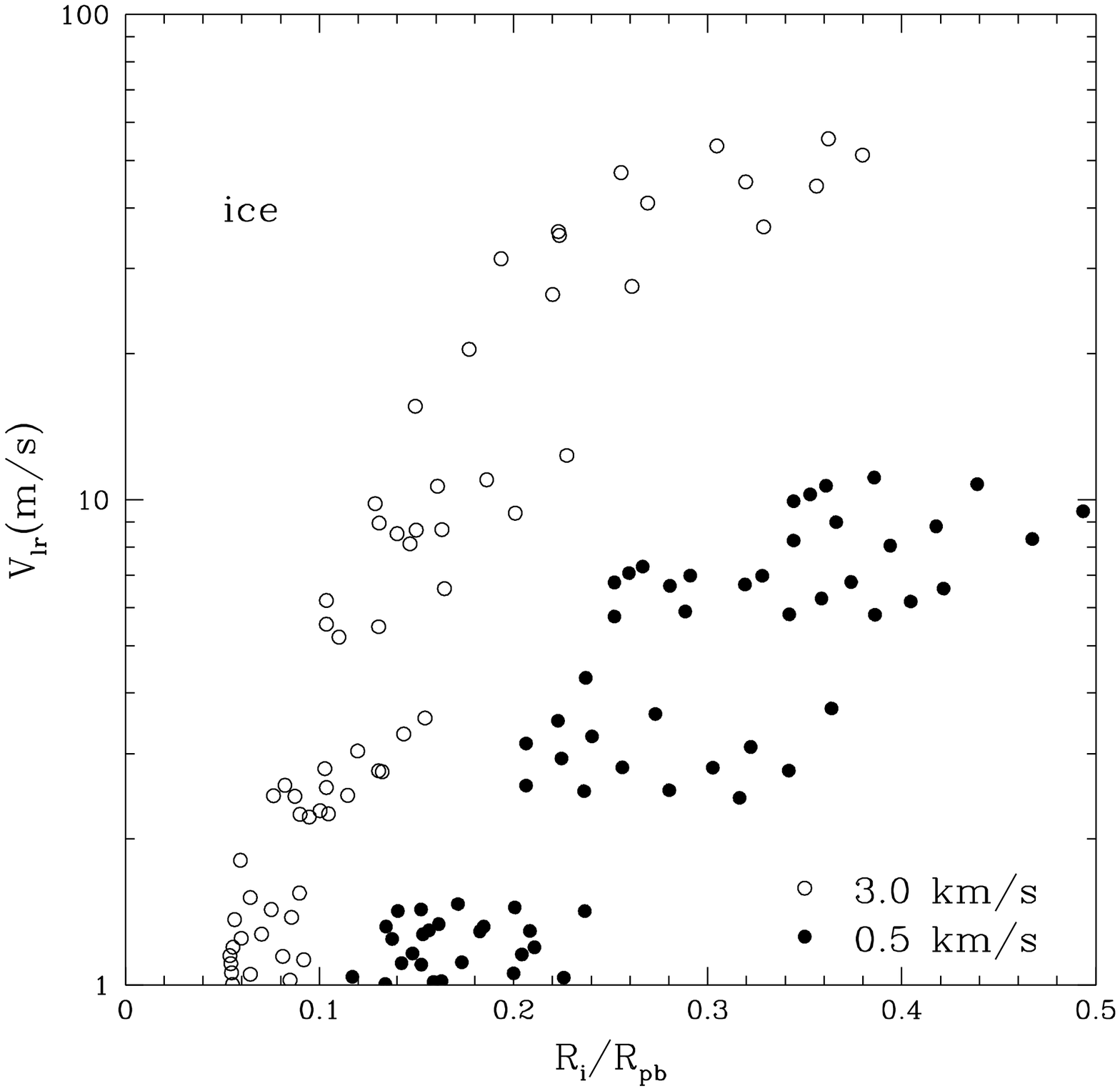,width=12cm}
  \caption{\ }
  \end{flushleft}
\end{figure}

\end{document}